\documentclass[twocolumn,amsmath,amssymb,superscriptaddress,floatfix,nofootinbib]{revtex4-1}

\usepackage{color}
\usepackage[colorlinks=true,linkcolor=blue,urlcolor=blue,citecolor=blue]{hyperref}
\usepackage{times}

\usepackage{tikz}
\usetikzlibrary{shapes,arrows}

\usepackage{graphicx}
\usepackage[margin=.7in,top=1in,bottom=1in]{geometry}
\usepackage{xspace}
\usepackage{pgfplots}
\usepackage[Algorithm]{algorithm}
\usepackage[noend]{algpseudocode}
\usepackage{mathtools}

\newcommand{\avg}[1]{\left \langle #1 \right \rangle}

\newcommand{\ie}{\textit{i}.\textit{e}.\xspace}
\newcommand{\eg}{\textit{e}.\textit{g}.\xspace}

\newcommand{\cO}{\mathcal{O}}
\newcommand{\cX}{\mathcal{X}}
\newcommand{\cT}{\mathcal{T}}

\newcommand{\tl}{t_{\rm l}}
\newcommand{\tm}{t_{\rm m}}
\newcommand{\tr}{t_{\rm r}}
\newcommand{\be}{\begin{equation}}
\newcommand{\ee}{\end{equation}}
\newcommand{\bal}{\begin{align}}
\newcommand{\eal}{\end{align}}

\begin{document}
\title{Sampling first-passage times of fractional Brownian Motion using adaptive bisections}
\author{Benjamin Walter} 
\affiliation{Department of Mathematics, Imperial College London, London SW7 2AZ, United Kingdom}
\author{Kay J\"{o}rg Wiese}
\affiliation{\mbox{Laboratoire de Physique de l'\'Ecole Normale Sup\'erieure, ENS, Universit\'e PSL, CNRS, Sorbonne} 
\mbox{Universit\'e, Universit\'e Paris-Diderot, Sorbonne Paris Cit\'e, 24 rue Lhomond, 75005 Paris, France.}}

\date{\today}
\begin{abstract}
We present an algorithm to efficiently sample first-passage times for fractional Brownian motion. 
To increase the resolution, an initial coarse lattice is successively refined  close to the target, by adding exactly sampled midpoints, where the probability that they reach the target is non-negligible.
	Compared to a path of $N$ equally spaced points, the algorithm achieves the same numerical accuracy $N_{\rm eff}$, while sampling only a small fraction of all points. Though this induces a statistical error, the latter is bounded for each bridge, allowing us to bound the total error rate  by a number of our choice, say $P_{\rm error}^{\rm tot}=10^{-6}$. 	
	This leads to   significant improvements in both memory and speed.
	For $H=0.33$ and $N_{\rm eff}=2^{32}$, we need $5\,000$ times less CPU time and $10\, 000$ times less memory than the classical Davies Harte algorithm. The gain  grows for $H=0.25$ and $N_{\rm eff} = 2^{42}$ to     $3\cdot 10^{5}$ for CPU and $10^6$ for  memory.
We estimate our algorithmic complexity as ${\cal C}^{\rm ABSec}(N_{\rm eff}) = {\cal O}\left(\left( \ln N_{\rm eff}\right)^{3}\right)$, to be compared to Davies Harte which has complexity ${\cal C}^{\rm DH}(N) = {\cal O}\left(N   \ln N  \right)$. Decreasing $P_{\rm error}^{\rm tot}$   results in a small increase in complexity, proportional to $\ln (1/P_{\rm error}^{\rm tot})$.
Our current implementation  is limited to  the values of $N_{\rm eff}$  given above, due to a loss
of floating-point precision. 
Our algorithm can   be adapted  to other extreme events and arbitrary Gaussian processes. It enables one to numerically  validate theoretical predictions that were hitherto inaccessible.
\end{abstract}

\maketitle

\tikzstyle{block} = [rectangle, draw, 
text width=9em, text centered, rounded corners, minimum height=5em]
\tikzstyle{state} = [diamond, draw, text width=4.5em, text badly centered, node distance=3cm, inner sep=0pt]

\section{Introduction}
Estimating the distribution of first-passage times  is a key problem in understanding systems as different as financial markets or  biological systems \cite{Redner2001,Metzler2014}, and the dynamics of local reactions \cite{Haenggi1990,GodecMetzler2016}. 
Typically, research   focuses on non-Markovian processes and bounded geometries, where first-passage time distributions are   difficult to obtain analytically \cite{Sanders2012,Guerin2016,Levernier2018,Wiese2018,ArutkinWalterWiese2019}.
Within the class of non-Markovian processes, fractional Brownian Motion (fBm) is of particular interest as it naturally extends standard diffusion to sub- and super-diffusive self-similar processes \cite{MandelbrotVanNess1968}.
Fractional Brownian Motion  is a one-parameter family of Gaussian processes, indexed by the Hurst parameter $H\in (0,1]$. 
The latter parametrizes the mean-square displacement via 
\be
\avg{X_t^2} = 2t^{2H}\ ,
\ee recovering standard Brownian Motion at $H=\frac12$. It retains from Brownian motion scale and translational invariance, both in space and time.

\begin{figure}
\centerline{\fboxsep0mm
\mbox{\setlength{\unitlength}{1cm}\begin{picture}(8.65,5.9)
\put(0,0){\includegraphics[width=\columnwidth]{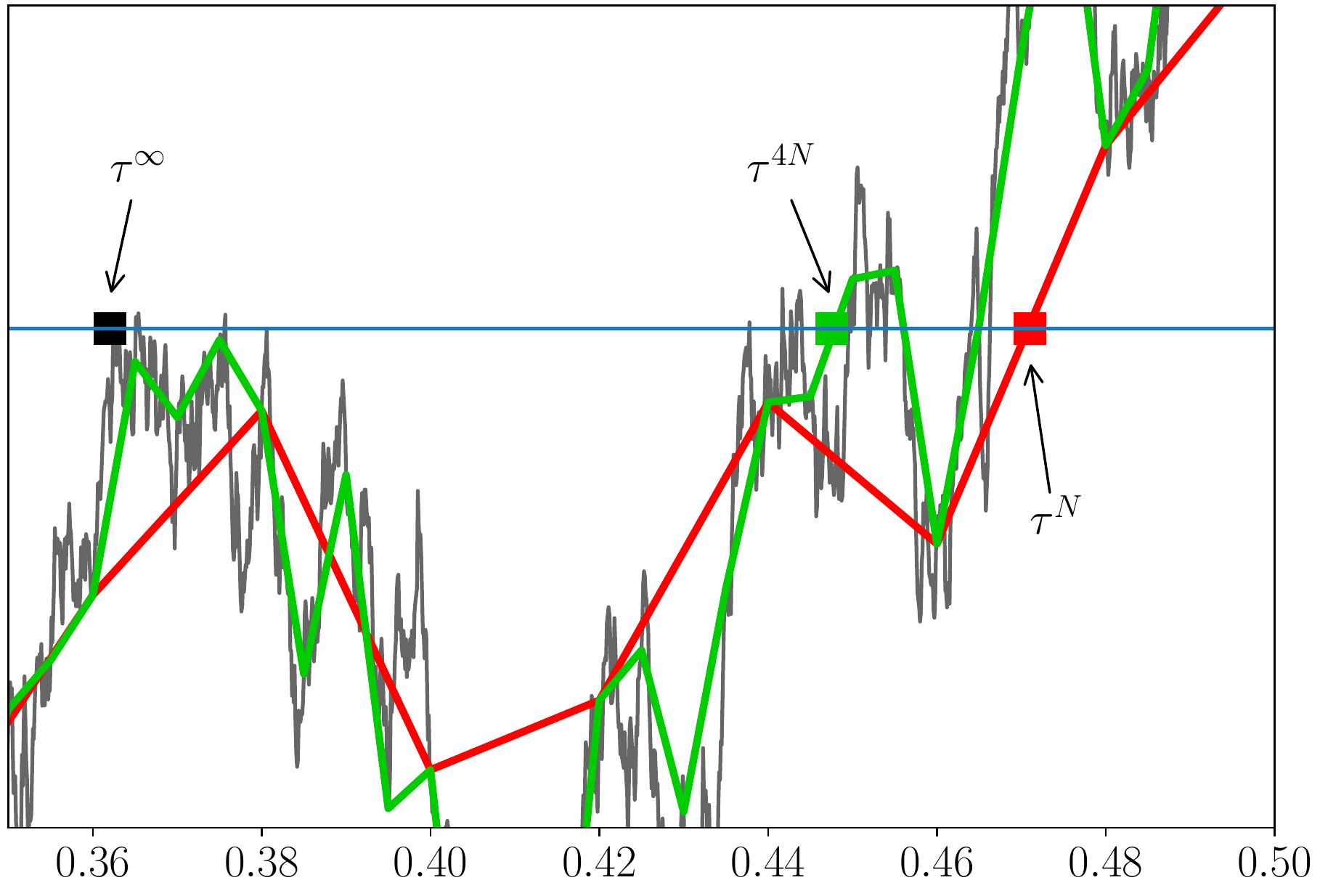}}
\put(8.2,0.6){$t$}
\put(0.2,5.5){$X_{t}$}
\end{picture}}}
\caption{The continuous stochastic path (grey) crosses the barrier (blue) for the first time at $\tau^{\infty}$ (black square mark). The discretization with $N$ points (red) over-estimates this time as $\tau^N$ (red square). The numerical estimate is improved to $\tau^{4N}$ (green square) when refining the discretization (green). This systematic error worsens for diminishing values of Hurst parameter $H$.}
\label{fig:discrete_error}
\end{figure}

In order to render the extreme events of this process accessible to an analytical treatment, an $\varepsilon$-expansion  around Brownian motion in $\varepsilon=H-\frac12$ has been proposed \cite{WieseMajumdarRosso2010}. This field theoretic approach has been applied to a variety of extreme events, yielding the first-order corrections of several probability distributions \cite{WieseMajumdarRosso2010,DelormeWiese2015, SadhuDelormeWiese2017, Wiese2018}.
The scaling functions predicted by this perturbative field theory   are computationally expensive to verify, since they   require a   high numerical resolution of the path. Typically this is done by simulating a discretized version of the path over a grid of $N$ equidistant points. Measuring a first-passage time then amounts to finding the first passage of a linear interpolation of these grid points.
This approximation, however, can  lead to a systematic over-estimation of the first-passage time. As can be seen on Fig.~\ref{fig:discrete_error}, a high resolution of the path is necessary in order to find the first-passage event at $t=0.36$ instead of     the one at $t=0.45$ or even $t=0.47$ for the coarser grids. 
To account for this, usually the number of grid points is increased. As the size of fluctuations  between gridpoints   diminishes as 
\be\label{delta-X}
\delta X = N^{-H}\ ,  
\ee
the sub-diffusive regime ($H<\frac12$) necessitates an enormous computational effort.

 As can be seen in Ref.~\cite{Wiese2018}, this poses challenges to the numerical validation of high-precision analytical predictions. As an example, for $H=0.33$, one needs system sizes of at least $N=2^{24}$, necessitating a CPU time of 6 seconds per sample.  
If theories of such high   precision are to be tested against simulations, new numerical techniques need to be developed. The present work addresses this problem by designing, implementing, and benchmarking a new algorithm sampling first-passage times of fractional Brownian Motion using several orders of magnitude less CPU time and memory than traditional methods. The general idea is to start from a rather coarse grid (as the red one on Fig.~\ref{fig:discrete_error}), and to refine the grid   where necessary. As a testing ground, we simulate and compare to theory the first-passage time of an fBm with drift \cite{ArutkinWalterWiese2019}.

The algorithm proposed  here is an adaptive bisection routine that draws on several numerical methods already established in this field, notably the Davies-Harte algorithm \cite{DaviesHarte1987}, bisection methods \cite{Ceperley1995,Sprik1985}, and the Random Midpoint Displacement method \cite{Fournier1982,Norros2004}. The central, and quite simple, observation is that in order to resolve a first-passage event it is necessary to have a high grid resolution only near   the target. This translates into an algorithm that generates a successively refined grid, where refinement takes place only at points close to the target, with the criterion of {\em closeness} scaling down by $2^{{-H}}$ for each bisection. This refinement is stopped after the desired resolution is reached. The sampling method is \emph{exact}, \ie the collection of points is drawn from the ensemble of fBm, a continuous process,  with no bias.
The only error one can make is that one misses an intermediate point. We have been able to control this error with a failure rate smaller than $10^{-6}$ per realisation.

 While there is a relatively large overhead for the non-homogenous refinement, this is compensated by the   use of far less points, leading  to a significant increase both in speed and memory efficiency over sampling methods that produce points for the full grid. For $H=0.33$ and system size $N=2^{32}$, our algorithm  is  5000 times faster  than the Davies Harte Algorithm (DH), the fastest exact   sampler (cf.~\cite{DiekerPhD, Craigmile2003}) if all points are needed. It has computational complexity $\cO(N \log(N))$, which makes it the standard algorithm in most current works, see \eg \cite{Guerin2016, Levernier2018, Krapf2019},  with system size $N$ ranging from $2^{21}$ to $2^{24}$.
Our maximal grid size is limited by the precision of the floating point unit to $N_{\rm max} \approx 2^{11/H}$.

This paper is organised as follows. In Sec.~\ref{sec:algorithm}, we introduce our adaptive bisection algorithm. First, its higher-level structure is outlined and then each subroutine is   detailed. Possible generalisations to other extreme events or other Gaussian processes are discussed at the end of this section. In Sec.~\ref{sec:results}, we present our implementation of the adaptive bisection in \texttt{C}, which is freely available \cite{WalterWiese2019a}. We benchmark it against an implementation of the Davies Harte algorithm. We compare error rates, average number of bisections, CPU time and memory. Sec.~\ref{sec:summary} contains a summary of our findings.

\section{Algorithm\label{sec:algorithm}}
In this section, we introduce the adaptive bisection routine (ABSec). The central aim is to translate the idea of refining the grid ``where it matters'' into a rigorous routine. 
\subsection{Fractional Brownian Motion and first-passage times}

Gaussian processes $X_t$ are stochastic processes for which $X_{t}$  evaluated at a finite number of points $  {\cal T}$ in  time, has a   multivariate Gaussian distribution \cite{Piterbarg2015}. They are  simple to handle, since their path probability measure can be obtained from their correlation function. The best known Gaussian Process is Brownian Motion which is the only translational invariant  Gaussian process with stationary and independent increments. 

Fractional Brownian Motion (fBm) generalises Brownian Motion by relaxing the requirement of independent increments, while keeping self-similarity. The latter property means that its path probability measure is invariant under a space-time transformation $t \to ct, \ x \to c^{-H}x$ for $c >0$. The parameter $H$ is referred to as \emph{Hurst} exponent.  As a Gaussian process, fBm is entirely characterized by its mean $X_0 = \avg{X_t} = 0$, and correlation function
\begin{align}
C(s,t) = \avg{X_s X_t} = |s|^{2H}+|t|^{2H}-|t-s|^{2H} \ ,
\label{eq:correlation_function}
\end{align}
where $H \in (0,1]$. As a consequence, $\avg{(X_{t}-X_{s})^{2}} = 2 |t-s|^{{2H}}$, and in particular 
 $\avg{X_{t}^{2}} = 2 |t|^{{2H}}$.
 From the correlation function it follows that on all time scales non-overlapping increments are positively correlated for $H > \frac12$ and negatively correlated for $H <\frac12$. For $H = \frac12$ one recovers Brownian Motion with uncorrelated increments. 

 The first-passage time (FPT) of a stochastic process is the fist time the process   crosses a threshold $m$. Since we use $X_0=0$, it is defined for $m>0$ as
 \begin{align}
 \tau_m = \inf_{t>0} \left\{ t | X_t \ge m \right\}.
 \label{}
 \end{align}

\subsection{Notation}
In simulating a fBm on a computer, one is forced to represent the continuous path by a \emph{discretized path} that takes values on a finite set of points in time, the \emph{grid}. We denote the grid by {\em ordered} times $\cT = \left \lbrace t_1, t_2, \cdots, t_N \right\rbrace$, and the corresponding values of the process by $\cX = \left \lbrace X_{t_1}, X_{t_2}, \cdots, X_{t_N} \right \rbrace$. Together, $(\cX, \cT)$ form the discretized path. 
Due to self-similarity of the process, we can restrict ourselves to $\cT \subset [0,1]$ with no loss of generality.
The intervals between any two successive times $t_i, t_{i+1} \in \cT$ are referred to as \emph{bridges} $(t_i, t_{i+1})$. Each connected component of $[0,1]\backslash \cT$ is a bridge. 

We denote the dyadic lattice on the unit interval by $\Lambda^k = \left\{ i\cdot 2^{-k} ; 0 \leq i \leq 2^k \right\}$. Our adaptive bisection algorithm sets out from a dyadic lattice $\cT^{(0)} = \Lambda^g$ of relatively low resolution (typically $g \lesssim 8$ or 10) and performs several bisections of that grid in successive iterations $\cT^{(0)}, \cT^{(1)}, \cdots, \cT^{(M)}$, where $M$ is the number of bisections generated before the routine terminates. To each bridge $(\tl, \tr)$ spanned in between left and right endpoints $\tl$ and $\tr$ and contained in a grid $\cT^{(m)}$, one can associate a level $\ell$ defined by $\ell = -\log_2 \left( \tr - \tl \right)$. 
A bridge is bisected by introducing its midpoint, $\tm = \frac12\left( \tl + \tr \right) = \tl + 2^{-\ell-1}$ and inserting it into the grid $\cT^{(m+1)} = \left\lbrace t_1, ..., \tl, \tm, \tr, ..., t_N \right\rbrace $. A bridge is allowed to be bisected until the level of a bridge reaches a \emph{maximum bisection level} $L$ (typically $L \lesssim 30$ for $H=0.33$). Since each iteration only   halves an existing interval, all grids are sandwiched between two dyadic lattices
\begin{align}
\Lambda^g = \cT^{(0)} \subset \cT^{(1)} \subset \cdots \subset \cT^{(M)} \subseteq \Lambda^L,
\label{}
\end{align}
representing the lowest and highest possible resolution. We define the truncation of the grid $\cT$ to a certain time $\tau \in [0,1]$ as
\begin{align}
{\cal T}|_\tau := \left\{ t_i \in \cT | t_{i-1}<\tau \right\}
\label{eq:truncation_grid}
\end{align}
\ie the truncation contains all points in time up to time $\tau$ \emph{and to the next highest gridpoint contained in the initial grid $\Lambda^g$} (cf.~section \ref{subsec:truncategrid}).
Note that for each bridge $(\tl, \tr)$, there is always one dyadic lattice $\Lambda^n$, s.t.\ $t_i$ and $ t_{i+1}$ are neighbouring points in $\Lambda^n$; they are members, but not neighbours in  $\Lambda^{n'}$ for $n'>n$, and at least one of them does not exist in  $\Lambda^{n'}$ for $n'<n$.

\subsection{Definition of the algorithm}
 The algorithm consists of two phases. In the {\em first phase}, the initialisation, a coarse grid is generated. In the {\em second phase}, the adaptive bisection, this grid is successively bisected   where necessary. 
 Once the second phase terminates, 
 the first-passage time is calculated using the final grid.

 The first phase starts by sampling an initial discretized path $\cX^{(0)}$ over a dyadic lattice $\cT^{(0)} = \Lambda^g$ with $N=2^g$ equidistant points,  using the \emph{Davies-Harte} salgorithm. The latter is the fastest known algorithm to sample an exact fBm path on an equidistant grid in time \cite{DiekerPhD}: its execution time scales as $N \ln (N)$, thus only slightly slower than what is needed to generate an uncorrelated sample of the same length $N$. 
 From this relatively coarse grid, $(\cX^{(0)}, \cT^{(0)})$, the first-passage time is estimated via   linear interpolation as $\tau^{(0)}$.
  
  Subsequently,  the grid is truncated by discarding   all points behind the first point surpassing $m$ (cf.~Eq.~\eqref{eq:truncation_grid}). 
That this does not change the measure is explained in section \ref{subsec:truncategrid}. 
  If no such point exists, the full grid is kept. The correlations between the different points $X_t$ at times $t$ stored in the grid are  given by the correlation matrix
\begin{align}
C_{ij}(\cT) = C(t_i, t_j) \ ,\qquad t_i, t_j \in \cT, 1 \leq i,j \leq |\cT| \ .
\label{eq:correlation_matrix}
\end{align}
It  is a symmetric matrix computed from the correlation function \eqref{eq:correlation_function}. It is then inverted to obtain the  inverse correlation matrix $C^{-1}_{ij}(\cT)$.  The inversion is optimised by using  pre-calculated tabularized matrices. 
This concludes the {\em first  phase}.

In the {\em second phase}, bridges are checked successively until the maximum level is reached. The order in which the bridges of the growing grid are checked is determined by a   subroutine whose aim   is to find the first-passage event with the least amount of bisections. The check consists in testing whether the  \emph{midpoint} $X_{\tm}$ of the bridge $(\tl, \tr)$  could surpass the threshold $m$ with a probability larger than $\varepsilon$, taken  small.
If this is the case  
the bridge is deemed \emph{critical} and bisected. The bisection consists in generating a midpoint $X_{\tm}$ at time $\tm$ conditional to the pre-existing grid. This computation requires the inverse correlation matrix and is detailed in Sec.~\ref{subsec:generatingmidpoint}. Once the   midpoint is generated, it is added to the path $(\cX, \cT)$. In a last step the inverse correlation matrix of the new grid, $C^{-1}(\cT \cup \tm)$  is  stored.
Further below, the algorithm is given in pseudocode.
\begin{figure}[t] 
\leftline{\noindent\rule{8.7cm}{0.6pt}}
\leftline{\bf ALGORITHM 1: Adaptive bisection}
\vspace*{-1ex}
\leftline{\rule{8.7cm}{0.4pt}}

\begin{algorithmic}
\Procedure{ABSec}{$g, L, m, \varepsilon$}
\State $\cT \gets \Lambda^g$
\State $\cX \gets  $\Call{davies\_harte}{$\Lambda^g$} \Comment{\ref{subsec:DHA}}
\State $\tau \gets $ \Call{fpt\_from\_grid}{$\cX, \cT$} \Comment{\ref{subsec:estimatefpt}}
\State $(\cX, \cT) \gets \left.(\cX, \cT) \right|_{\tau^{(0)}}$ \Comment{\ref{subsec:truncategrid}}
\State $C^{-1} \gets$ \texttt{CMatrixTable}[$\tau^{(0)}$] \Comment{\ref{subsec:Tabulating}}
\State $(\tl, \tr) \gets $\Call{Next bridge}{$\cT$, 0, $\tau^{(0)}$} 
\While{$(\tl, \tr)$ defined}
\If{Bridge $(\tl, \tr)$ critical and not yet bisected} \Comment{\ref{subsec:bridgecritical}}
\State $C^{-1} \gets \Call{Augment ${C^{-1}}$-Matrix}{C^{-1}, \tm }$
\State $X^* \gets \Call{Generate midpoint}{C^{-1}, \tm } $ \Comment{\ref{subsec:generatingmidpoint}}
\State $\cX \gets \cX \cup X^*$
\State $\cT \gets \cT \cup \tm$
\If{$X^*>m$}
\State $\tau \gets \Call{fpt\_from\_grid}{\cX, \cT}$
\EndIf
\EndIf
\State $(t_i, t_{i+1}) \gets $\Call{Next bridge}{$\cT$, $(t_i, t_{i+1})$, $\tau$} \Comment{\ref{subsec:bridgeselection}}
\EndWhile
\State \textbf{output}({$\tau$})
\EndProcedure
\end{algorithmic}
\label{algo:bsec}
\vspace*{-1ex}
\leftline{\rule{8.7cm}{0.6pt}}
\end{figure}
The routines   in the pseudocode are described   in   sections \ref{subsec:DHA}--\ref{subsec:bridgeselection}.

\subsubsection{Davies-Harte Algorithm}
\label{subsec:DHA}

The Davies-Harte algorithm (DH) is a widely used method to generate fBm samples. It was    introduced in \cite{DaviesHarte1987}, is pedagogically described in \cite{DiekerPhD}, and has been extended  to other Gaussian processes in \cite{Craigmile2003}, allowing us to   omit an introduction. It    generates a sample of fractional Gaussian noise (fGn) $\xi_1, \xi_2, \cdots, \xi_N$, the incremental process of fBm $\xi_j = X_{j+1} - X_j, j \in \mathbb{N}$, and   then sums the increments to a fBm sample with values $X_{i \cdot \delta t} = (\delta t)^{H} \sum_{j=1}^{i} \xi_j$. 
Simulating the increments is more efficient since fGn is a stationary Gaussian process which, for equally sized increments, has a circulant correlation matrix, which can be diagonalised  using a fast Fourier transform (FFT). Therefore a fGn sample of $N$ increments can be simulated with computational complexity $\cO(N \log(N))$. The FFT algorithm works optimally when the number of points is a natural power of $2$, \ie if the grid is a dyadic lattice. 

\subsubsection{Estimating the first-passage time}
\label{subsec:estimatefpt}
Given a discretized path $(\cX, \cT)$, we use its linear interpolation to give the first-passage time as its first intersection with the threshold (cf.~Fig.\ref{fig:discrete_error}).

\subsubsection{Truncating the grid}
\label{subsec:truncategrid}
A further  optimisation is to discard  grid points  beyond the first point   crossing the threshold (cf.~Eq.~\eqref{eq:truncation_grid}). It is necessary to show that the density of first-passage times conditioned on the full grid equals the distribution conditioned on the truncated grid, \ie that truncating does not change the measure. 

The first-passage time distribution (FPTD) $P(\tau)$ can be decomposed into a sum of conditional probabilities for disjoint events: Each term of the sum is the probability that the $i$th point of a grid surpasses $m$, the threshold, \emph{for the first time} (``$P^{\rm grid}(X_{t_i} > m \text{ first})$''), times the FPTD  of a fBm conditioned on the event that its discretization on grid $\cT$ surpasses $m$ at $t_i$ for the first time, \ie
\begin{multline}
 P_{\cT}(\tau |  X_{t_i} > m \text{ first}) \\
 = P(\tau | X_t: X_{t_i} > m \text{ and } X_{t_j} < m \, \forall  t_j < t_i) 
\end{multline}
for $t_j, t_i \in \cT$.
The decomposition thus reads
\begin{align}
P(\tau) = \sum_{t_i \in \cT}^{} P_{\cT}(\tau |  X_{t_i} > m \text{ first}) P^{\rm grid}(X_{t_i} > m \text{ first}) .
\label{eq:decompositionofprobabilities}
\end{align}
By continuity of the process,
\begin{align}
P_{\cT}(\tau > t_i | X_{t_i} > m \text{ first}) = 0 ,
\label{eq:continuity}
\end{align}
such that the sum in
 Eq.~\eqref{eq:decompositionofprobabilities} can be truncated to
 \begin{align}
P(\tau) = \sum_{t_{i-1} < \tau}^{} P_{\cT}(\tau |  X_{t_i} > m \text{ first}) P^{\rm grid}(X_{t_i} > m \text{ first}) .
\end{align}
In order to sample $P_{\cT}(\tau |  X_{t_i} > m \text{ first})$, one would naively sample the entire grid $\cX$ over all of $\cT$, but since
\be
P_{\cT}(\tau  | X_{t_i} > m \text{ first}) = P_{\cT  |_{ \tau}}(\tau | X_{t_i} > m \text{ first}),
\ee 
where the restriction is defined in Eq.~\eqref{eq:truncation_grid},  it is sufficient to only regard the smaller grid $\cT|_{\tau}$, \ie
\begin{align}
P(\tau) = \sum_{t_{i-1} < \tau}^{} P_{\cT|_{\tau}}(\tau |  X_{t_i} > m \text{ first}) P^{\rm grid}(X_{t_i} > m \text{ first}) .
\end{align}
Discarding   points in the initial stage leads to a smaller  correlation matrix to be inverted, which  increases performance, and decreases memory.

 \subsubsection{Tabulating inverse correlation matrices}
 \label{subsec:Tabulating}
 The inverse of the correlation matrix \eqref{eq:correlation_matrix} is necessary to compute the conditional probability of  any further midpoint    (cf.~App.~\ref{app:derivationmidpoint}). Its computation is costly and typically scales with $\mathcal{O}(N^3)$ where $N=2^g$ is the number of points in $\cT^{(0)}$. If the algorithm is run multiple times,  this computation   slows it down. The initial grid however, is always   a dyadic lattice truncated at some point, \ie $\cT^{(0)} = \left\{ k2^{-g}; 0 \leq k \leq K \right\}$, where $X_{K2^{-g}}$ is the first point to surpass $m$.
  Therefore, the initial inverse correlation matrix $C^{-1}\left( \cT^{(0)} \right)$ can   take  $2^g-1$ possible values, one for each possible value of $K$. It is more efficient to pre-calculate   all possible inverse correlation matrices in the beginning, and store them in a vector `\texttt{CMatrixTable}',
\begin{align}
\mathtt{CMatrixTable}[K] = \left(\left[ C(i\cdot 2^{-g}, j\cdot 2^{-g})\right]_{i,j=1}^{K}  \right)^{-1}\ .
\label{eq:CMatrixTable}
\end{align}
After generating the initial grid and measuring $\tau^{(0)}$, one reads out the appropriate entry of the table at $K = \min \left\lbrace n \in \mathbb{Z}; n2^{-g} \geq \tau^{(0)}\right \rbrace$.

\subsubsection{Deciding whether a bridge is critical}
\label{subsec:bridgecritical}
Once entering the bisection phase, the algorithm needs to decide whether a particular bridge is \emph{critical}, \ie whether it is suspicious  of hiding a ``dangereous'' excursion crossing the threshold at $m$ (cf.~Fig.~\ref{fig:discrete_error}). Rather than determining whether \emph{any} point in $(\tl, \tr)$ surpasses the threshold, we focus on the midpoint $\tm$ conditioned on all other points $\cX$, and ask how likely $X_{\tm} > m$. Such an event needs to be avoided with a very low probability $\varepsilon$, the \emph{error tolerance}. The relevant probability,
\be\label{11}
P\left(X_{\tm} > m | \cX \right)< \epsilon ,
\ee
is too costly to be computed for every bridge in every step of the iteration, as  the midpoint is a Gaussian random variable, with its mean and variance determined by every other point in the grid. If we   \emph{ignore} all points of the path apart from $(t_i, X_{t_i})$ and $(t_{i+1}, X_{t_{i+1}})$, a calculation given in App.~\ref{app:criticalstrip} shows that mean and variance would be given by 
\be \mu = \frac12\left( X_{t_i} + X_{t_{i+1}} \right)
\ee
and 
\be\sigma^2 = \left( 2^{1-2H} - \frac12 \right)\times 2^{-2 \ell H} \ . 
\label{eq:sigmanoneighbour}
\ee
Here $\ell$ is the level of the bridge of width $\delta t = 2^{-\ell}$. Interestingly,  adding to the  bridge's endpoints further points {\em lowers} the variance (cf.~Eq.~\eqref{eq:sigma_from_grid})  which means that neglecting all but nearest neighbours gives an upper bound on the variance of the midpoint. Further, we conservatively bound the mean by the maximum of both endpoints, $\mu \lesssim \max\left( X_{t_i}, X_{t_{i+1}} \right)$. This is a priori not a precise approximation, since far-away grid points are  able to ``push'' the expected midpoint   above the bridges' endpoints for values of $H\neq \frac12$. As is shown   in Sec.~\ref{subsec:errortolerance}, this systematic error can be absorbed by introducing an even smaller error tolerance   $\varepsilon'$. Furthermore, it is less relevant in the sub-diffusive regime, where the process is negatively correlated. 
 By giving conservative bounds on mean and variance with quantities that are {\em local} (\ie do not depend on the remaining grid), we can replace the original criterion  (\ref{11})  by a computationally   cheaper alternative, namely the   local condition 
\be
P\left( X_{\tm} > m |  (X_{\tl}, X_{\tr})\right)< \epsilon' .
\label{eq:localcriterion}
\ee
This implies that Eq.~(\ref{11}) holds for an appropriate choice of $\varepsilon'$, on {\em average}. 
This is to be understood as follows: In a simulation, there are $n$ decisions of type \eqref{11} to be taken. The total error is  approximately 
$
P_{\rm error}^{\rm tot } \approx n \epsilon
$. 
The parameter $\epsilon'$ is chosen such that the total error rate remains smaller than $10^{-6}$, and thus negligible as compared to  MC fluctuations. 
The dependence between   $\epsilon'$ and $P_{\rm error}^{\rm tot }$ is   investigated in Sec.~\ref{subsec:errortolerance} (cf.~Fig.~\ref{fig:errorgrowth}).

Criterion \eqref{eq:localcriterion} is rephrased, using again $\ell$ as the level of the bridge, to
\begin{align}
\Phi\left( \frac{m - \max\left( X_{t_i}, X_{t_{i+1}}\right) }{\left(\sqrt{2^{1-2H }-\frac12}\right)2^{-\ell H}} \right) > 1 - \varepsilon' 
\end{align}
which implies
\begin{align}
\max\left( X_{t_i}, X_{t_{i+1}}  \right) < m - \left(\sqrt{2^{1-2H}{-}\frac12}\right)2^{-\ell H} \Phi^{-1}(1{-}\varepsilon') ,
\label{eq:criticalstripcondition}
\end{align}
where we introduced $\Phi$, the cumulative distribution function of the standard normal distribution, and its inverse $\Phi^{-1}$. This is further simplified by defining the \emph{critical strip} 
\begin{align}
c_0 = \left( \sqrt{2^{1-2H}-\frac12} \right) \Phi^{-1}\left( 1- \varepsilon' \right)\ ,
\label{eq:criticalstrip}
\end{align}
and the level-corresponding critical strips
\begin{align}
c_{\ell} = 2^{-\ell H} c_0.
\label{}
\end{align}
A bridge $(X_{\tl}, X_{\tr})$ of level $\ell$ is   deemed critical if either of its endpoints lies above the critical strip corresponding to $\ell$, \ie
\begin{align}
\max\left( X_{t_i}, X_{t_{i+1}}  \right) > m - c_{\ell} \ .
\label{eq:localcriterion2}
\end{align}
This makes for a computationally fast decision process, since the critical strip width has to be computed only once. The procedure   then   checks for a given level of the bridge  whether it reaches into the critical strip, in which case it is bisected (cf.~Fig.~\ref{fig:illustrationalgo} for illustration).

\subsubsection{Generating the new midpoint efficiently}
\label{subsec:generatingmidpoint}
If a bridge triggers a bisection, the midpoint is drawn according to its probability distribution, given all   points that  have  been determined previously. If this occurs  at, say, the $m$-th iteration, the discretized path is $((X_1, t_1),  \cdots, (X_N, t_N))$ with $|\cT^{(m)}| = |\cX^{(m)}| = N = K+m$ where $K\leq 2^g$ is the number of points in the truncated initial grid. Denoting the midpoint to be inserted by $(X_{t^*}, t^*)$, one needs to find
\begin{align}
P\left( X_{t^*} | X_1, \cdots, X_N \right).
\label{}
\end{align}
The midpoint is again  normal distributed with  mean $\mu_*(N)$ and variance $\sigma_*(N)$. Let $\cT^{(m+1)} = \left(\cT^{(m)}, t^* \right)$ be the augmented grid, and $C^{-1}(N+1) = C^{-1}\left( \cT^{(m+1)} \right)$  the associated inverse correlation matrix (cf.~Eq.~\eqref{eq:correlation_matrix}). Then, as detailed in App.~\ref{app:derivationmidpoint}, the inverse of the variance is given by
\begin{align}
 \sigma^{-2}_*(N)  =  \left[C^{-1}(N+1)\right]_{N+1, N+1} \ ,
\label{eq:conditionalsigma}
\end{align}
and the mean by
\begin{align}
\mu_*(N) = \sigma^2_*(N) \sum_{i=1}^{N} \left[ C^{-1}(N+1) \right]_{N+1, i} X_{t_i}.
\label{eq:conditionalmu}
\end{align}
Computing the inverse correlation matrix from scratch at every iteration  would require a   matrix inversion   which typically uses $\mathcal{O} ( N^3  )$ steps. 
We do this in $\cO ( N^2  )$ steps, by starting from the already calculated inverse 
correlation matrix of the previous grid $C^{-1}(N) = C^{-1}\left( \cT^{(m)}  \right)$. As  detailed in App.~\ref{app:derivationaugmentedmatrix}, the inverse correlation matrix  
$C^{-1}(N+1)=C^{-1}\left( \cT^{(m+1)}  \right)$ can be constructed as follows: First,  generate a vector containing all correlations of the new point with the grid, using Eq.~\eqref{eq:correlation_function}
\begin{align}
\vec{\gamma}(N) = \left( C(t^*, t_1), C(t^*, t_2), \cdots, C(t^*, t_N) \right)^T.
\label{eq:defgammavec}
\end{align}
Second,   multiply it with the (already constructed) inverse correlation matrix to obtain
\begin{align}
\vec{g}(N) = C^{-1}(N) \vec{\gamma}(N)\ .
\label{eq:defgvec}
\end{align}
In terms of $\vec{\gamma}$ and $\vec{g}$, the mean and variance can be expressed as
\begin{align}
\mu_*(N) = \vec{X}^T \vec{g},
\label{eq:mu_from_grid}
\end{align}
where we use $\vec{X} = \left( X_{t_1}, \cdots, X_{t_N} \right)$ for short, and
\begin{align}
\sigma^2_*(N) = 2 \left( t^* \right)^{2H} - \vec{\gamma}^T \vec{g}.
\label{eq:sigma_from_grid}
\end{align}
Since $\vec{\gamma}^T \vec{g} = \vec{\gamma}^T C^{-1}(N) \vec{\gamma} > 0$, conditioning on more points  
diminishes the variance of a midpoint. 
The outer product of $\vec{g}$ defines the matrix
\begin{align}
G(N) := \vec{g} \otimes \vec{g}^T\ .
\label{eq:defGmatrix}
\end{align}
It is used to build the enlarged inverse correlation matrix
\begin{align}
C^{-1}(N+1) = \begin{pmatrix}
C^{-1}(N) + \sigma^{-2} G(N) &\vline & -\sigma^{-2} \vec{g}(N)\\
\hline 
-\sigma^{-2} \vec{g}^T(N) & \vline & \sigma^{-2}
\end{pmatrix}\ ,
\label{eq:cmatrixpromoted}
\end{align}
where   $\sigma^2 = \sigma^2_*(N)$.

\subsubsection{Bridge selection}
\label{subsec:bridgeselection}
The task of the bridge-selection routine (cf.~Alg.~\ref{algo:bsec}) is to choose the order in which bridges of the successively refined grid are tested, and possibly inserted. Its aim is to find the first-passage event with the least number of bisections. To this aim,  it zooms in into areas where a first-passage time is likely, and zooms out when the possibility of a crossing becomes negligible. In this subsection, we phrase this intuition in more rigorous terms.

Prior to the first call of the routine, the initial grid consists of $2^g$ bridges of uniform width $2^{-g}$. The routine selects the earliest bridge, \ie $(\tl=0, \tr=2^{-g})$, and scans all bridges of the initial grid in ascending order in time until a critical bridge is found (by applying the criticality criterion \eqref{eq:localcriterion}).  Once such a bridge is found, the algorithm explores this   bridge by successive bisections. After a finite number of bisections the algorithm either has identified a first-passage event to the desired precision, or no crossing was found. In the latter case, the routine then moves on to the next bridge of the initial grid.

 In order to illustrate the workings of the bridge-selection routine, it is helpful to consider a bijection between the adaptively bisected grid and a rooted binary tree (cf.~Fig.~\ref{fig:illustrationalgo}). Every bridge $(\tl, \tr)$ that is bisected by introducing a point at $\tm$ contains two sub-bridges $(\tl, \tm)$ and $(\tm, \tr)$. We refer to these bridges as the \emph{left} and \emph{right} children of $(\tl, \tr)$. Vice versa, every bridge that is not part of the initial bridge (\ie with level $\ell > g$) is the child of another bridge which is referred to as parent of the bridge. The set of all bridges that are contained in a initial bridge of width $2^{-g}$ is mapped to a rooted binary tree by identifying every node with a bridge, where a node can either have zero or two children depending on whether the bridge has been bisected or not. The root of the tree corresponds to the bridge contained in the initial bridge from where the bisections were spawned off. The generation of a node in the tree corresponds to its level by $\text{generation} = \ell -g +1$. Therefore, the depth of the tree is limited to $\text{generation}_{\rm max} = L - g + 1$.
 
 The routine stores a representation of this tree internally, together with the information whether a node/bridge has previously been  checked for criticality or not.  If a bridge is bisected, but its two children have not yet been checked for criticality, the left child is selected. This is because earlier crossings of the threshold render later crossings irrelevant. If a bridge has two children, but the left has already been checked  (implying that neither it nor any of its further descendants contains a first-passage event), the right child is selected. If both children of the bridge have already been checked,  none of the descendants contains a first-passage event. In that case the parent of the bridge is returned (zooming out). If the routine returns to the root, the bridge of type $(i \cdot 2^{-g}, (i+1) \cdot 2^{-g})$ has no  parent, and the next such bridge $((i+1)2^{-g}, (i+2)\cdot 2^{-g})$ is returned. If $i=2^{g}-1$, the routine terminates by returning an empty bridge since the entire grid has been checked.

To summarise, the  routine is either descending (zooming in) or ascending (zooming out) within the tree, depending on whether the children of a node, if existent, have been visited or not. 

The routine takes into account two additional constraints. First,  the maximum bisection level $L$; if a bridge of maximum level $L$ contains a first-passage event, the routine terminates since this estimate has reached the desired resolution. If it contains no crossing, the parent is returned. Second, it takes into account whether a bridge is early enough in time to improve the first-passage estimate. If a bridge at level $\ell$ records a first-passage event, only its descendants can improve this result.

We give the pseudocode of the routine below. In the implementation we present later (Sec.~\ref{subsec:implementation}), the algorithm is implemented slightly differently for performance reasons. The logical steps however are the same and we decided to present them here for pedagogical reasons.
\begin{figure}[t]
\leftline{\noindent\rule{8.7cm}{0.6pt}}
\leftline{\bf ALGORITHM 2: {Finding the next bridge to be checked}}
\vspace*{-1ex}
\leftline{\rule{8.7cm}{0.4pt}}
\begin{algorithmic}
\Procedure{Next Bridge}{$\cT, (\tl, \tr), \tau_m$}
\If {$(\tl, \tr)=\texttt{0}$}
\State \textbf{return} $(0,2^{-g})$ \Comment{Initialise with first bridge}
\EndIf
\If{$(\tl, \tr)$ has no children}
\State \textbf{return} parent bridge
\EndIf
\If{$(\tl, \tr)$ early enough for $\tau_m$ \texttt{AND} level $<L$ } 
\If{left child not checked}
\State  \textbf{return} left child \Comment{Move down left}
\EndIf
\If{left child checked \texttt{AND} right child not checked}
\State  \textbf{return} right child \Comment{Move down right}
\EndIf
\If{both children  checked}
\State  \textbf{return} parent bridge \Comment{Move up to parent}
\EndIf
\EndIf
\If{level of $(\tl, \tr)$ = $L$}
\If{Bridge crosses threshold}
\State \textbf{return} \texttt{NULL}
\Else 
\State \textbf{return} parent bridge
\EndIf
\EndIf
\EndProcedure
\end{algorithmic}
\vspace*{-1ex}
\leftline{\rule{8.7cm}{0.6pt}}
\end{figure}

\tikzset{myptr/.style={decoration={markings,mark=at position 1 with     {\arrow[scale=3,>=stealth]{>}}},postaction={decorate}}} 
\begin{figure}
\centerline{
\fboxsep0mm
\mbox{~~~~\resizebox{1.06\columnwidth}{!}{
\begin{tikzpicture}
\node at (0,-0.5) {$i \cdot 2^{-g}$};
\node at (8,-0.5) {$\!\!\!\!(i+1) \cdot 2^{-g}~~~~$};
\node (n1) at (0,0) {$\bullet$};
\node (n2) at (4,0) {$\bullet$};
\node (n3) at (5,0) {$\bullet$};
\node (n4) at (5.5,0) {$\bullet$};
\node (n5) at (6,0) {$\bullet$};
\node (n6) at (8,0) {$\bullet$};
\node[circle,draw=black,inner sep=0mm,minimum size=2mm, fill=blue] (1) at (4,7) {};
\node[circle,draw=black,inner sep=0mm,minimum size=2mm]  (2) at (2,6.5) {};
\node[circle,draw=black,inner sep=0mm,minimum size=2mm, fill=blue]  (3) at (6,6.5) {};
\node[circle,draw=black,inner sep=0mm,minimum size=2mm, fill=blue]  (4) at (5,6) {};
\node[circle,draw=black,inner sep=0mm,minimum size=2mm]  (5) at (7,6) {};
\node[circle,draw=black,inner sep=0mm,minimum size=2mm]  (6) at (4.5,5.5){}; 
\node[circle,draw=black,inner sep=0mm,minimum size=2mm, fill=blue]  (7) at (5.5,5.5){};
\node[circle,draw=black,inner sep=0mm,minimum size=2mm, fill=red]  (8) at (5.25,5) {};
\node[circle,draw=black,inner sep=0mm,minimum size=2mm]  (9) at (5.75,5) {};
\draw (1) node[above] {1}-- node[above, pos=1.05] {2} (2);
\draw (1) -- (3);
\draw (3) node[above] {3} -- (4);
\draw (3) -- (5);
\draw (4) node[above] {4} -- node[above, pos=1.05] {5} (6);
\draw (4) -- node[above, pos=1] {6} (7);
\draw (7) -- node[above, pos=1] {7} (8);
\draw (7) -- (9);
\draw (-1,0) node[above] {$t$} --  (8.5,0);
\draw[gray, thick] (0,1) -- (8,3);
\draw[gray,thick] (0,1) -- (4,1.3) -- (8,3);
\draw[gray,thick] (0,1) -- (4,1.3) -- (6,2.6) -- (8,3);
\draw[gray,thick] (0,1) -- (4,1.3) -- (5, 2.1) -- (6,2.6) -- (8,3);
\draw[very thick] (0,1) -- (4,1.3) -- (5, 2.1) -- (5.5, 2.7) -- (6,2.6) -- (8,3);
\draw[thick, red] (0,2.5) node[above] {$m$} -- (8,2.5);
\draw[blue] (0,1.1) node[left] {$m-c_g$} -- (8,1.1);
\draw[blue] (0,1.9) node[left] {$m-c_{g+1}$} -- (8,1.9);
\draw[blue] (0,2.35) node[left] {$m-c_{g+2}$} -- (8,2.35);
\draw[dashed] (n1) -- (0,1);
\draw[dashed] (n2) -- (4,1.3);
\draw[dashed] (n3) -- (5,2.1);
\draw[dashed] (n4) -- (5.5,2.7);
\draw[dashed] (n5) -- (6,2.6);
\draw[dashed] (n6) -- (8,3);
\draw[thick] (.5,7)-- (.5,4.75);
\node () at (-.2,7) {$g$};
\node () at (-.2,6.5) {$g+1$};
\node () at (-.2,5) {$L$};
\node () at (-2,4.) [rotate=90] {Bridges};
\node () at (-2,6.) [rotate=90] {Binary tree};
\node () at (-2,1.5) [rotate=90] {Path $\cX$};
\node () at (-2,0) [rotate=90] {Grid $\cT$};
\draw[<->] (0,4.5)-- (8,4.5);
\draw[<->] (0,4.25)-- (4,4.25);
\draw[<->] (4,4.25)-- (8,4.25);
\draw[<->] (4,4.)-- (6,4.);
\draw[<->] (6,4.)-- (8,4.);
\draw[<->] (4,3.75)-- (5,3.75);
\draw[<->] (5,3.75)-- (6,3.75);
\draw[<->] (5,3.5) -- (5.5,3.5);
\draw[<->] (5.5,3.5) -- (6,3.5);
\draw[dotted, thick] (1) -- (4,4.5);
\draw[dotted, thick] (2) -- (2,4.25);
\draw[dotted, thick] (3) -- (6,4.25);
\draw[dotted, thick] (4) -- (5,4.0);
\draw[dotted, thick] (5) -- (7,4);
\draw[dotted, thick] (6) -- (4.5,3.75);
\draw[dotted, thick] (7) -- (5.5,3.75);
\draw[dotted, thick] (8) -- (5.25,3.5);
\draw[dotted, thick] (9) -- (5.75,3.5);
\end{tikzpicture}
}}}
\caption{Illustration of the adaptive bisection routine. The grid $\cT$ (bottom) contains points in time, here detail shown of initial bridge $\tl = i2^{-g}, \tr=(i+1)2^{-g}$ (labelled bullets) and successively introduced midpoints (bullets on time axis); The path $\cX$ (above) samples values at times (dashed lines) which approximate path by linear interpolations (grey and black thick lines). The threshold $m$ (red line) is crossed by the path and bisections are generated for every bridge whose endpoints lie in the critical strip corresponding to its level (blue lines). The horizontal arrows on top of the path indicate the bridges in between the grid points. The mapping from bridges to binary tree (top) is indicated with dotted lines. The top node (1) corresponds to the widest bridge $(i2^{-g}, (i+1)2^{-g})$, and children correspond to sub-intervals generated by midpoint. The bridges are explored in order as given by numbers above nodes and chosen by the bridge-selection routine (see text for details). Bridges that are critical (blue nodes) are bisected, and their children checked from left to right, until a first-passage event has been identified at maximum bisection level $L$ (red node). This event terminates the algorithm.}
\label{fig:illustrationalgo}
\end{figure}
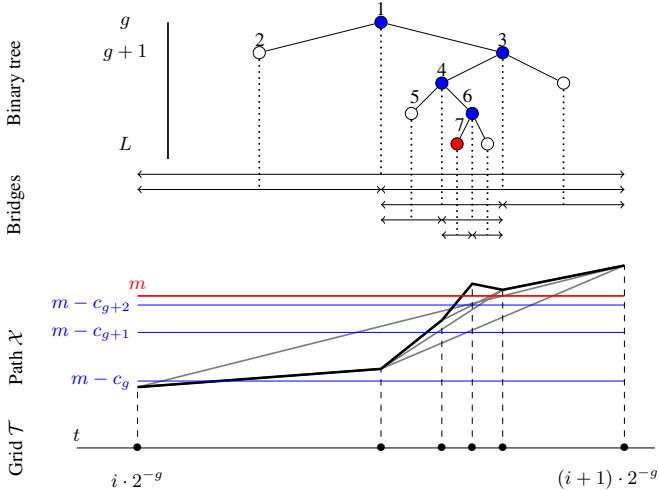

\subsection{Adding deterministic functions}
The adaptive bisection routine can be adapted to further generate first-passage times of stochastic processes of the form
\begin{align}
Z_t = X_t + f(t),
\end{align}
where $f(t)$ is a deterministic smooth function, \eg a linear or fractional drift term, and $X_t$ is again a fractional Brownian motion. In its first phase, $X_t$ is generated on a subgrid, and $f(t)$ is added accordingly. The resulting process $Z_t, t \in \cT^{(0)}$ is then passed to the bridge-selection routine, where the bridges are checked for criticality using the values of $Z_t$ in the criticality criterion \eqref{eq:criticalstripcondition}. Once a bisection is required, the midpoint is generated using the subtracted process $X_t = Z_t - f(t)$, \ie the vector used to generate the midpoint's mean (cf.~Eq.~\eqref{eq:conditionalmu}) is $\vec{X}$, not $\vec{Z}$. Then, the generated midpoint $X_{\rm m}$ is transformed back using $Z_{\rm m} = X_{\rm m} + f(t_{\rm m})$, and inserted into the path of $Z$. 
Note that even if $f(t) = \mu t$ (linear drift), and contrary to Brownian motion, the iteration can not be performed directly on $Z_{t}$.

\subsection{Further  generalisations}
The underlying idea of the algorithm -- to generate a grid that is fine only where it matters -- lends itself to various other non-local observables, in particular extreme events, such as running maxima (minima), positive time (time spent in the region $X_t>0$), last returns, or the range or span ($\max X_t - \min X_t$) of a process. 

In each of these examples, one   needs to adapt two logical steps in the procedure: $(i)$ the order in which bridges of the grid are iterated, and $(ii)$ the criterion for triggering a bisection. For first-passage times, the order of the bridges is given by the subroutine described above in Sec.~\ref{subsec:bridgeselection}. The criterion for bisection is determined by   the bridge's distance to the threshold. These two choices are particular to first-passage events. 

For running maxima, the bridges should be tested in descending order of height, and the bisection-criterion adapted to decide whether the midpoint could surpass the current maximum with a probability larger than $\varepsilon$. 
If the current maximum changes, the criterion for triggering a bisection also changes. As the maximum can only increase, bridges which were uncritical before do not become critical by a change of the estimate of the maximum.

To find the  last return to zero ($t_0 = \sup_{t'<t} \left\{ t' | X_{t'} = 0 \right\}$), the bisection criterion is the same as for first-passage times (with $m$ set to zero), but bridges should  be iterated over from latest to earliest, choosing the right subinterval first after bisection (cf.~Fig.~\ref{fig:illustrationalgo}). 

The span of a process at time $t$ is defined as the running maximum minus the running minimum \cite{WeissRubin1976,WeissDiMarzioGaylord1986,PalleschiTorquati1989,Wiese2018,Wiese2019}. 
To find the first time the span reaches one  is more delicate. There are two cases, given a discretization: Either span one is reached first  when   the maximum increases, or the minimum decreases. Suppose that the maximum increases. Then there is   a minimum for a smaller time. By refining the grid close to this minimum, the latter may decrease. This in turn shifts down  the critical strip for the 
maximum, and one has to redo all checks for bridges close to the maximum. 

The algorithm can  be generalized to other Gaussian processes, since the derivations given in Sec.~\ref{subsec:generatingmidpoint} and App.~\ref{app:derivationmidpoint} for the insertion of a conditional midpoint apply to \emph{any} Gaussian process. The only point at which we made explicit use of properties for fBm was at the initialisation step, where the Davies-Harte method was employed to generate a path on a coarse dyadic lattice. If one were to study another Gaussian process, one would   need to replace the correlation function \eqref{eq:correlation_function}, and adapt the routine generating the initial grid.

Once these  modifications are made for the new problem,  we expect the  algorithm  to  deliver similar improvements in performance and memory.

\section{Results and Benchmarking}
\label{sec:results}
In this section, we compare an implementation of our adaptive bisection method (ABSec) with an implementation of the Davies Harte (DH) method. Our   focus lies on comparing both CPU time and memory usage for a simulation of equal discretization error. We find that for large system sizes, $N_{\rm eff} \gtrsim 10^{2/H}$, the adaptive bisection routine outperforms the Davies Harte method both in CPU time and memory. This advantage grows markedly for lower values of $H$. At $H=0.33$, for instance, and a final grid size of $N_{\rm eff}  = 2^{32}$ we need 5000 times less CPU time and 10 000 less memory. At $H=0.25$ we find ABSec to be 300.000 times faster and $  10^6$ less memory intensive than DH at an effective system size of $N_{\rm eff} = 2^{42}$.

We then discuss   systematic errors and analyse how they depend on the parameters, in order to clarify the payoff between computational cost and numerical accuracy.
 We conclude with a discussion of our findings.

\subsection{Implementation in \texttt{C}}
\label{subsec:implementation}
We implemented the adaptive biection algorithm in $C$, using external libraries \texttt{lapack} \cite{lapack}, \texttt{gsl} \cite{gsl}, \texttt{fftw3} \cite{Frigo1999}, and \texttt{cblas} \cite{Blas}. The code is published 
\cite{WalterWiese2019a}
and   available under a BSD license. It was compiled using the Clang/LLVM compiler using the $\mathtt{-O3}$ flag as only compiler\ optimisation. The code was executed on an `Intel(R) Core(TM) i5-7267U CPU \@ 3.10GHz' processor.

As reference, we use an implementation of the Davies-Harte method in $C$\footnote{B.\,Walter, K.\,J.\,Wiese, \url{https://github.com/benjamin-w/davies-harte-fpt.git}}. Compiler settings  and hardware are identical to those used for the adaptive bisection algorithm.

In order to compare performance, we used user time and maximum resident set size as measured by the POSIX command \texttt{getrusage}; user time indicates the time the process was executed in user space, and maximum resident set size the amount of RAM held by the process.

\subsection{Numerical errors and fluctuation resolution}
\label{subsec:errors}
The adaptive bisection algorithm suffers from three errors. 

({\em i})  the   resolution of the grid itself, determined by the maximum grid size if all bridges were triggered, which we refer to as \emph{horizontal error}. Any discretization of a continuous path suffers from errors that are made when replacing the rough continuous path by the linear interpolation of a grid. Even if the true first-passage time is optimally approximated, the error   still depends on the system size $N$. In that respect, our algorithm does not differ from DH or other exact sampling methods. 

({\em ii}) the adaptive bisection routine suffers from a probabilistic error, namely false negative results of the criticality check, \ie bridges which do contain an excursion crossing the threshold $m$, but whose endpoints do not lie in the critical strip (cf.~Sec.~\ref{subsec:bridgecritical}). We refer to these errors as \emph{vertical errors}.

({\em iii}) the algorithm suffers from rounding errors of the floating-point unit.

Horizontal errors correspond to the resolution of the process' fluctuations. To contain fluctuations of a fBm between two grid points at distance $N^{-1}$ to the order of $\delta X$, one needs to choose $N \sim (\delta X)^{-\frac1H}$. Horizontal errors are therefore characterised by the effective discretization resolution $N^{H}\sim (\delta X)^{-1}$ which corresponds to the inverse fluctuation resolution. In order to compare two discretizations of a fBm path for two different values of the Hurst parameter $H$, comparing   $N$ is misleading. Rather, we compare their effective discretization resolutions $N^{H}$. Horizontal errors are impossible to measure numerically, since there exists no way to simulate a continuous path. They are however independent of the sampling method used; this implies that the horizontal error of a path generated by DH with system size $2^L$ and an adaptive bisection routine of maximum bisection level $L$ are \emph{exactly} the same, given no vertical error occurred. For a deeper discussion of discretization errors of the DH algorithm, see \cite[Sec.V.E]{Wiese2019}.

Vertical errors are controlled by the error tolerance $\varepsilon'$, of~Eqs.~\eqref{eq:criticalstripcondition}-\eqref{eq:criticalstrip}. To study vertical errors systematically, one needs to compare the results with a fully sampled grid using (for instance)   DH. This is discussed in the next section.

In the remainder of the section, we run benchmarking experiments that repeat the adaptive bisection routine a large number of times, typically $I=10^4$. Following the insights of Sec.~\ref{subsec:errortolerance}, we choose an error tolerance that is small enough to neglect errors of the vertical kind (whenever the vertical error rate is much smaller than $I^{-1}$). In doing so, we can ignore the vertical error such that the numerical discretization error becomes a good common error   for both adaptive bisections and   DH. This allows us to compare grids sampled with both methods systematically across various values of $H$ and $L$.

Finally, errors due to the finite precision of the floating-point unit are   considered. These   arise   in the  matrix inversion 
\eqref{eq:cmatrixpromoted}, where inspection reveals terms of opposite sign.  They can be detected by plotting $\sigma^2_*(N)$ as a function of grid resolution. For small grids, $\sigma^2_*(N)$ almost follows a power-law, with little spread. Numerical errors are visible as a net increase of this spread, see Fig.~\ref{fig:variance_error}. To be on the safe side, we choose the maximal $L$ to be 4 less than the point where we first see numerical errors appear.

\subsection{Error rate depending on $\varepsilon'$}
\label{subsec:errortolerance}

\begin{figure}[t]
\centering
\centerline{\fboxsep0mm
\mbox{\setlength{\unitlength}{1cm}\begin{picture}(8.65,5.6)
\put(0,0){\includegraphics[trim=5 5 60 50,clip,width=\columnwidth]{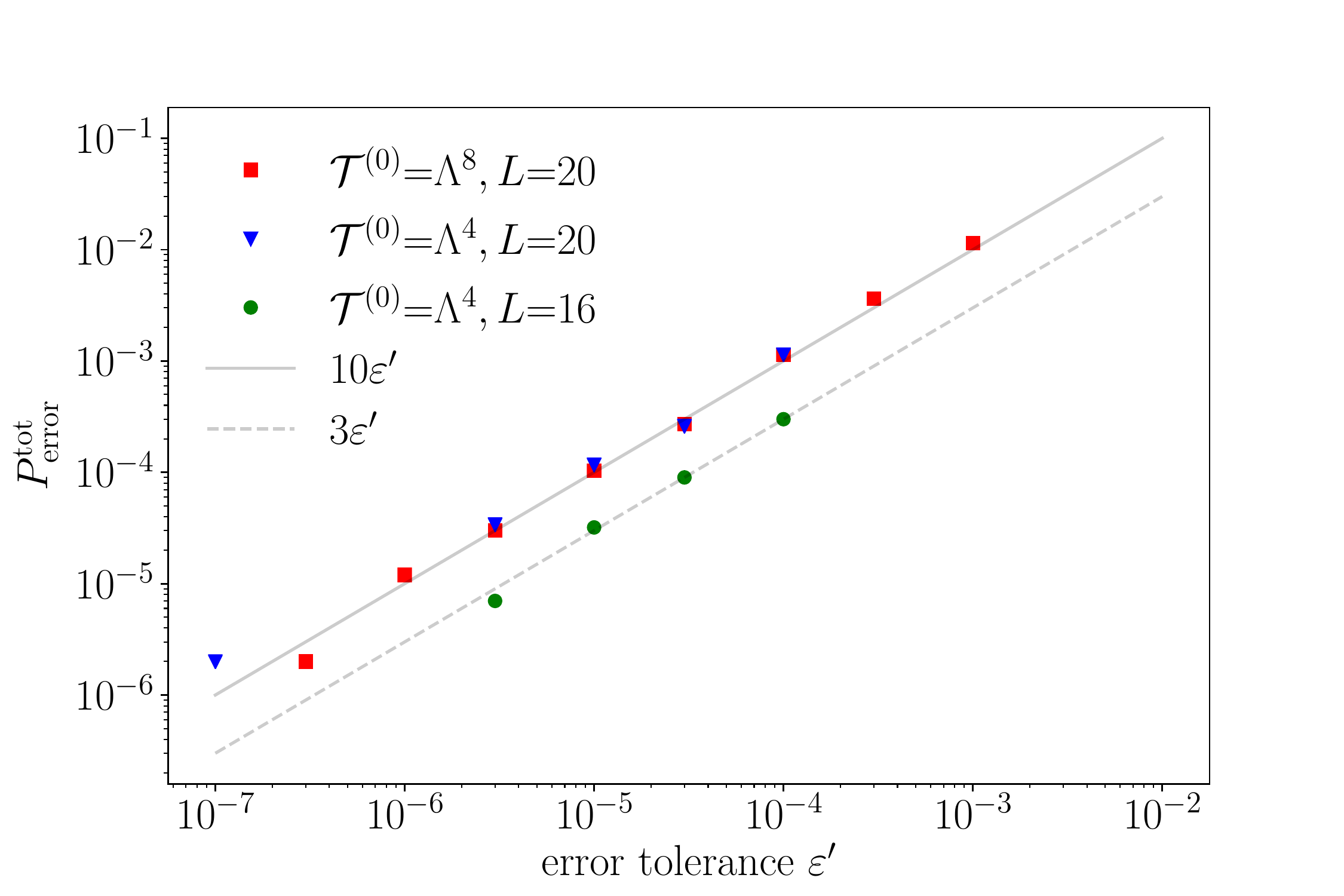}}
\end{picture}}}
\caption{Error rate from phone book test for various values of $\varepsilon'$ for $H=0.33$. The error rate is almost identical when changing the initial grid size from $2^8$ (red squares) to $2^4$ (blue triangles) at the same  maximum bisection level $L=20$. When lowering the maximum bisection level to $L=16$, the error rate improves.
 The relation between error rate and error tolerance decreases approximatively linearly over several orders of magnitude (compare with gray line). The total error rate is approximately $10   \varepsilon'$ for $L=20$ (solid gray line) and about $3 \varepsilon'$ for $L=16$ (dashed gray line). Note that the prefactor is much smaller than the number of points, which can read off from Fig.~\ref{fig:MvsL}.  Error rates were averaged over $10^5$ to $10^6$ iterations.}
\label{fig:errorgrowth}
\end{figure}

\begin{figure}[t]
\centering
\centerline{\fboxsep0mm
\mbox{\setlength{\unitlength}{1cm}\begin{picture}(8.65,5.75)
\put(0,0){\includegraphics[trim=28 10 60 50,clip,width=\columnwidth]{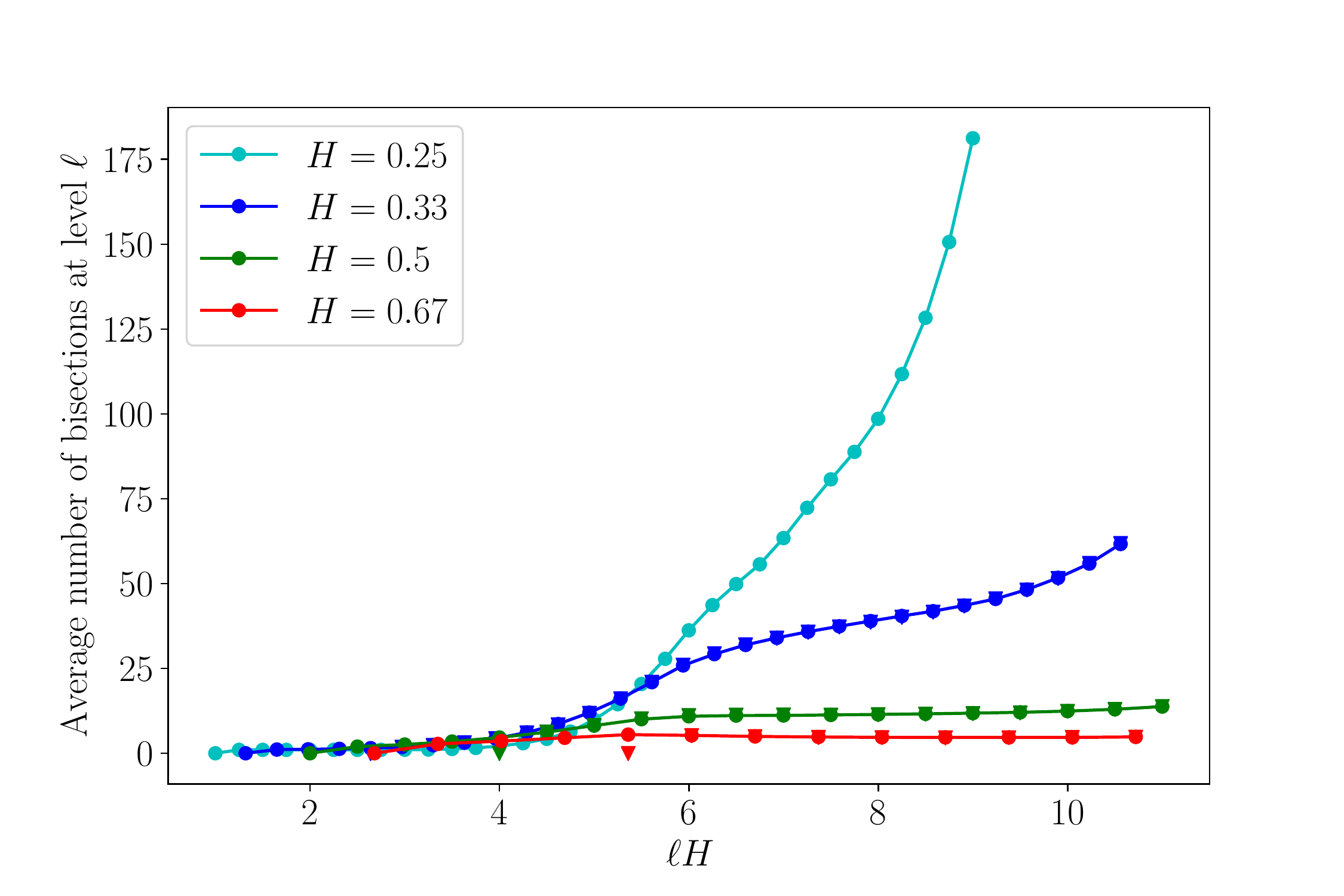}}
\end{picture}}}
\caption{Average number of new midpoints generated at   bridge level $\ell$, for various values of $H$  as a function of  $\ell H$. For equal values of $\ell H$, lower Hurst parameter implies a larger number of average bisections. 
These numbers are virtually independent of the  initial grid size, as shown for $\Lambda^{4}$ (circles) and $\Lambda^{8}$ (triangles).   
}
\label{fig:bisection_vs_H}
\end{figure}

This section addresses the question of vertical errors, \ie bridges that were deemed uncritical by the adaptive bisection routine (cf.~Sec.~\ref{subsec:bridgecritical}), yet contained an excursion that crossed the threshold for the first time.
This probability, $P(X_{\tm} > m)$, where $X_{\tm}$ marks the midpoint of a bridge, was bounded using an  error tolerance $\varepsilon'$. Therefore, we need to know how $\varepsilon'$ controls the error rate. Since we can only  measure the error rate when compared to another numerically generated grid, we compare our algorithm to a path generated using the Davies-Harte algorithm of equal precision. The procedure is as follows: In a first step, the Davies-Harte method is used to generate a path on the dyadic lattice $\Lambda^L$. For this path, and a threshold $m$, the first-passage time is calculated using its linear interpolation as detailed in Sec.~\ref{subsec:estimatefpt}. Then, only  times in the subgrid $\Lambda^g \subset \Lambda^L$ are copied into a second path. This path is handed over to a modified adaptive bisection routine (cf.~Alg.~\ref{algo:bsec}). The bridges of the grid are successively checked, at each step deciding whether to bisect as discussed in Sec.~\ref{subsec:bridgecritical}. Once a midpoint needs to be drawn,  it is not randomly generated, but taken from the  full grid at the same time. The full grid thus serves as a {\em phone book} for the adaptive bisection algorithm, where points are {\em looked up} if they lie at points the algorithm would have otherwise generated randomly. The algorithm then outputs its own estimate of the first-passage time. If the first-passage times disagree, this is considered an error. We refer to this check as \emph{phone book test}. This test is iterated  $10^6$ times, and the error rate $P_{{\rm error}}^{\rm tot}$ is defined as the ratio between errors and the number of iterations. 

The results are shown in Fig.~\ref{fig:errorgrowth}, where we compare the error-rate for different values of $\varepsilon'$ and for three different grids of varying initial  grid size, and maximum bisection levels. The plot shows that the total error rate and error tolerance  $\varepsilon'$ depend on each other  linearly, indicating that $\varepsilon'$ is a suitable replacement for $\epsilon$ introduced in Eq.~(\ref{11}). The plot further shows that the error rate remains almost identical when replacing the initial grid $\Lambda^8$ by $\Lambda^4$ (which contains 16 points only). Further, the error rate improves if the maximum bisection level is lowered. When lowering the effective system size from $2^{20}$ to $2^{16}$, the error rate lowers approximately by a factor of three. 

In summary, this plot confirms that the  computationally cheap variant \eqref{eq:localcriterion} allows us to control  the  vertical errors (false negative results of the criticality test).

\subsection{Average number of bisections}
\label{subsec:averagebisection}

\begin{figure}[]
{\fboxsep0mm
\mbox{\setlength{\unitlength}{1cm}\begin{picture}(8.7,5.68)
\put(0,0){{\includegraphics[trim=20 13 63 45,clip,width=1\columnwidth]{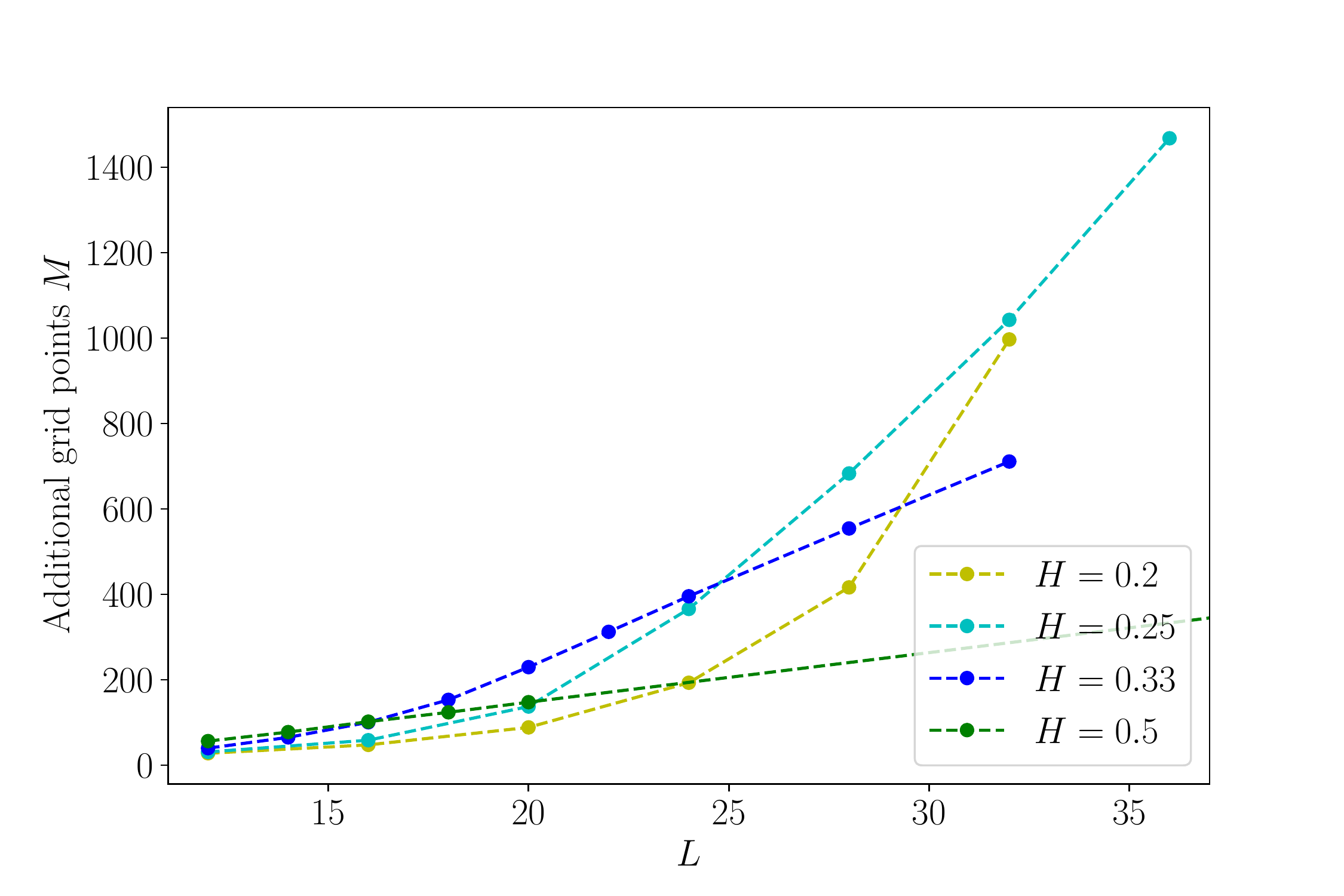}}}
\put(0.98,2.65){{\includegraphics[trim=38 38 37 32,clip,width=0.54\columnwidth]{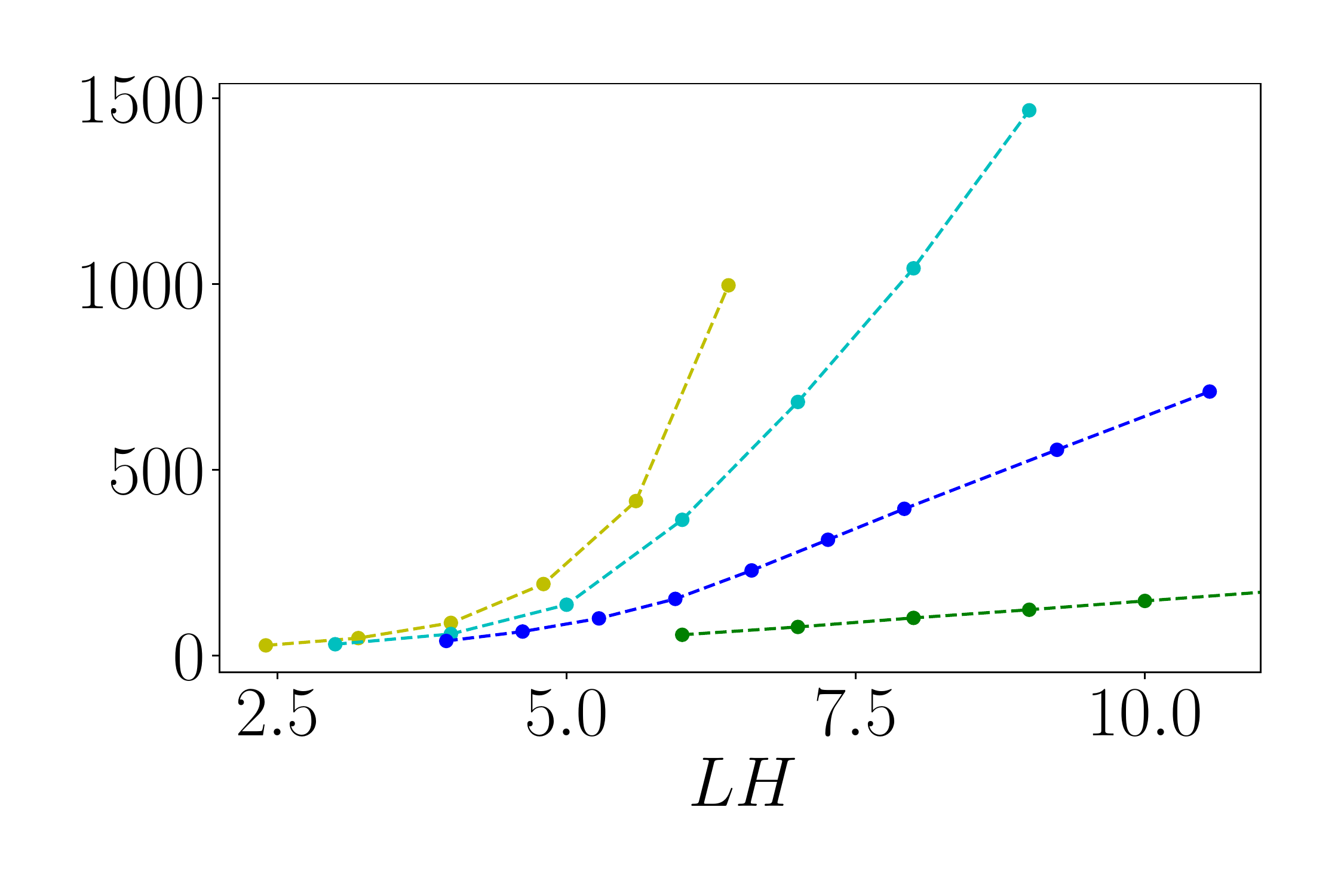}}}
\end{picture}}}
\caption{Average number of bisections $M$ as a function of the  maximum bisection level $L$ (i.e.\ $N_{\rm eff} = 2^{L}$) for different values of $H$ (Inset: $M$ versus $LH$). As long as $H\ge 0.33$  growth is asymptotically approximately linear in $L$, corroborating $M \sim \log(N_{\rm eff})$. 
For smaller values of $H$, either the linear regime is not yet reached, or the growth is stronger. 
(5000 iterations with initial grid $\Lambda^8$ and error tolerance $\varepsilon'=10^{-9}$).
For $H=0.5$ extrapolation was used.} 
\label{fig:MvsL}
\end{figure}
In this section, we investigate how many points are added to the initial grid, and how the additionally inserted midpoints are distributed over the different generations. The number of midpoints generated, $M$, is the main expense of computational resources, since each point requires promoting an inverse correlation matrix from size $n$ to  $n+1$ requiring $\cO(n^2)$ steps.

Each midpoint that is generated bisects a bridge at level $\ell$ and creates two   sub-bridges at level $\ell+1$. In order to know how the algorithm spends most of its time, we simulated the adaptive bisection routine $10^4$ times over an initial grid of size $\Lambda^4$ or $\Lambda^8$ and measured the average distribution of the $M$ newly generated midpoints over the different levels. The results are shown in Fig.~\ref{fig:bisection_vs_H}. 

While the distribution remains virtually unchanged when replacing the initial grid by $\Lambda^8$, its shape  changes for different values of Hurst parameter $H$. 
For $H>\frac12$, the distribution remains flat and even descends for $\ell > 5/H$. For $H=\frac12$ it remains constant for $\ell > 8$ (at around 11 midpoints per generation), while for $H<\frac12$ (see figure for $H=\frac13$ and $H=\frac14$), the number of inserted midpoints increases, and tends to be at higher bridges.

Since the number of additionally inserted points $M$ is crucial to the performance of ABSec, the routine is designed   to minimise this number, with a hypothetical minimum of $L - g$ points (when finding the first-passage event with no fault). The hypothetical maximum corresponds to a full bisection of the grid which would require $2^L - 2^g \approx 2^L$ additional points (this occurs when the path does not cross the threshold at all and $\varepsilon' \to 0$).  In Fig.~\ref{fig:MvsL}, we show the total number of bisections $M$ for various system sizes $L$, averaged over $10^4$ realisations. The number of additional points ranges from 40 to 1500, where larger system sizes lead to an increase of $M$. For $H=0.33$ and $L=32$, the average of additional points is $M=710$ which corresponds to $1.6 \times 10^{-7}$ of the full grid. This means that with that fraction of the full grid only, the algorithm identifies the first-passage time to the same accuracy as  DH (up to  vertical errors controlled by $\varepsilon'=10^{-9}$ in this case).

 We observe that for values of $H \gtrsim \frac13$, the number of bisections grows first sublinearly and then linearly in $L$. This behaviour changes for values $H \lesssim \frac14$, where growth is stronger, and we may not yet be in the asymptotic regime. This is also indicated by the profiles shown in Fig.~\ref{fig:bisection_vs_H}, where for lower values of $H$ the distribution ceases to tend to a plateau, but grows for higher levels of bisection $\ell$. 

\subsection{Computing time and complexity estimate}
\label{subsec:performancetest}
In this section, we analyse how the performance of our algorithm varies with different parameters, and how it compares to DH. In loose terms, we expect the initial grid, generated by DH, to cost $\cO(2^g \log(2^g))$, and each of the $M$ bisections to cost $k^2$ with $k$, the number of gridpoints, \ie costs, or more precisely the algorithmic complexity,  should behave as
\be
{\cal C}^{\rm ABSec}(g,M) \sim \sum_{k=2^g}^{2^g+M} k^2 \approx \frac13 (2^g+M)^3\ .
\label{eq:complexity_sum}
\ee 
It is therefore evident that the majority of the computational cost lies in the bisection phase, and the overall complexity is of order $\cO((2^g+M)^3)$. When comparing this to the complexity of generating $2^L$ gridpoints with   DH, which is $\cO(2^L \log(2^L))$, one   estimates that   ABSec   outperforms DH whenever $M^3 \ll 2^{L} \log(2^L)$. 
As is shown below, ABSec outperforms DH   for $L \gtrsim 12$ to $16$.

We define the performance of the algorithm via its user time, \ie the share of the CPU time the process spends in user space. This means that, depending on the implementation,  the total of CPU time (``wall time'')  might differ. User time  is a more robust observable, so we use it as best approximation to the   performance of the implementation.

\begin{figure}
\centering
\mbox{\includegraphics[trim=20 10 63 50,clip,width=1\columnwidth]{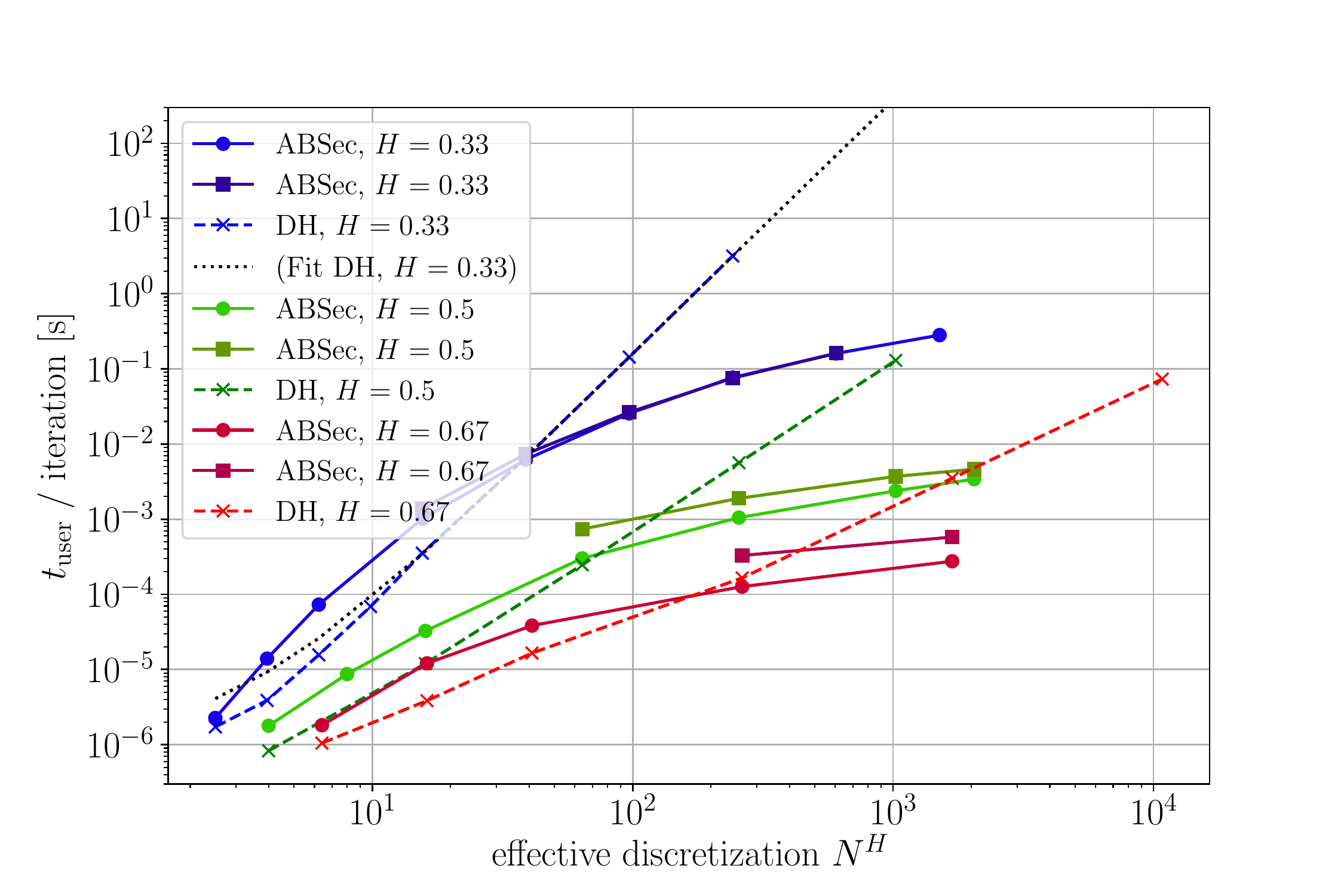}}
\caption{Average user time required to find first-passage time in a grid of effective discretization precision $2^{-LH}$. The dashed lines indicate user time for Davies-Harte method, solid lines for the adaptive bisection method. The three different colours indicate $H=0.33,0.5,0.67$. Simulations were run $10^4$ times for $\varepsilon' = 10^{-9}$ and for two different initial subgrid sizes ($\Lambda^4$ circles; $\Lambda^8$ squares).}\label{fig:performance}
\end{figure}
We measure the average user time per generated first-passage time,  using either   DH or ABSec. To render different Hurst-values and algorithms comparable, we plot the user time versus the inverse of the effective discretization error, which scales as $N^{H}$ for DH and $2^{LH}$ for ABSec. It describes how well the fBm-path is resolved numerically, taking into account the fluctuation scaling for different Hurst-parameters. 

Since at the beginning of the ABSec procedure inverse correlation matrices are tabulated (cf.~section \ref{subsec:Tabulating}), we measured the run time for $10^4$ iterations,  in order to render the initial overhead irrelevant.

Fig.~\ref{fig:performance} shows the result  of the benchmarking. For small effective system sizes,   ABSec performs slower than DH, which is due to the relatively complex overhead of bisections. For (effective) system sizes of $N \gtrsim 10^{\frac{2}{H}}$ the ABSec algorithm gains an increasing and significant advantage since its run time only grows sublinearly.

\begin{figure}
\centering
\centerline{\fboxsep0mm
\mbox{\setlength{\unitlength}{1cm}\begin{picture}(8.65,5.8)
\put(0,0){\includegraphics[trim=23 12 60 35,clip,width=1\columnwidth]{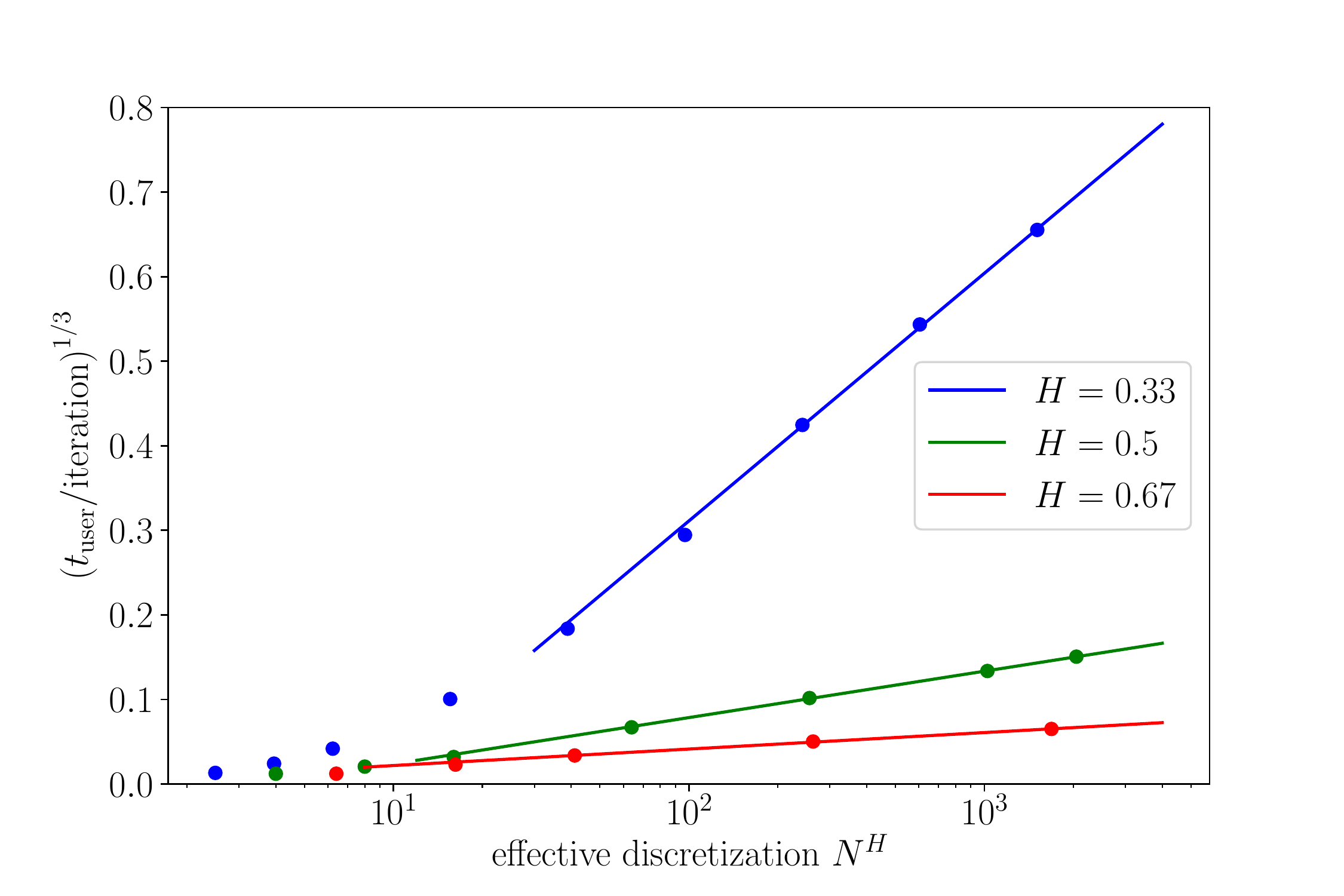}}
\put(1.1,3.1){\mbox{\includegraphics[trim = 30 25 40 55, clip, width=0.47\columnwidth]{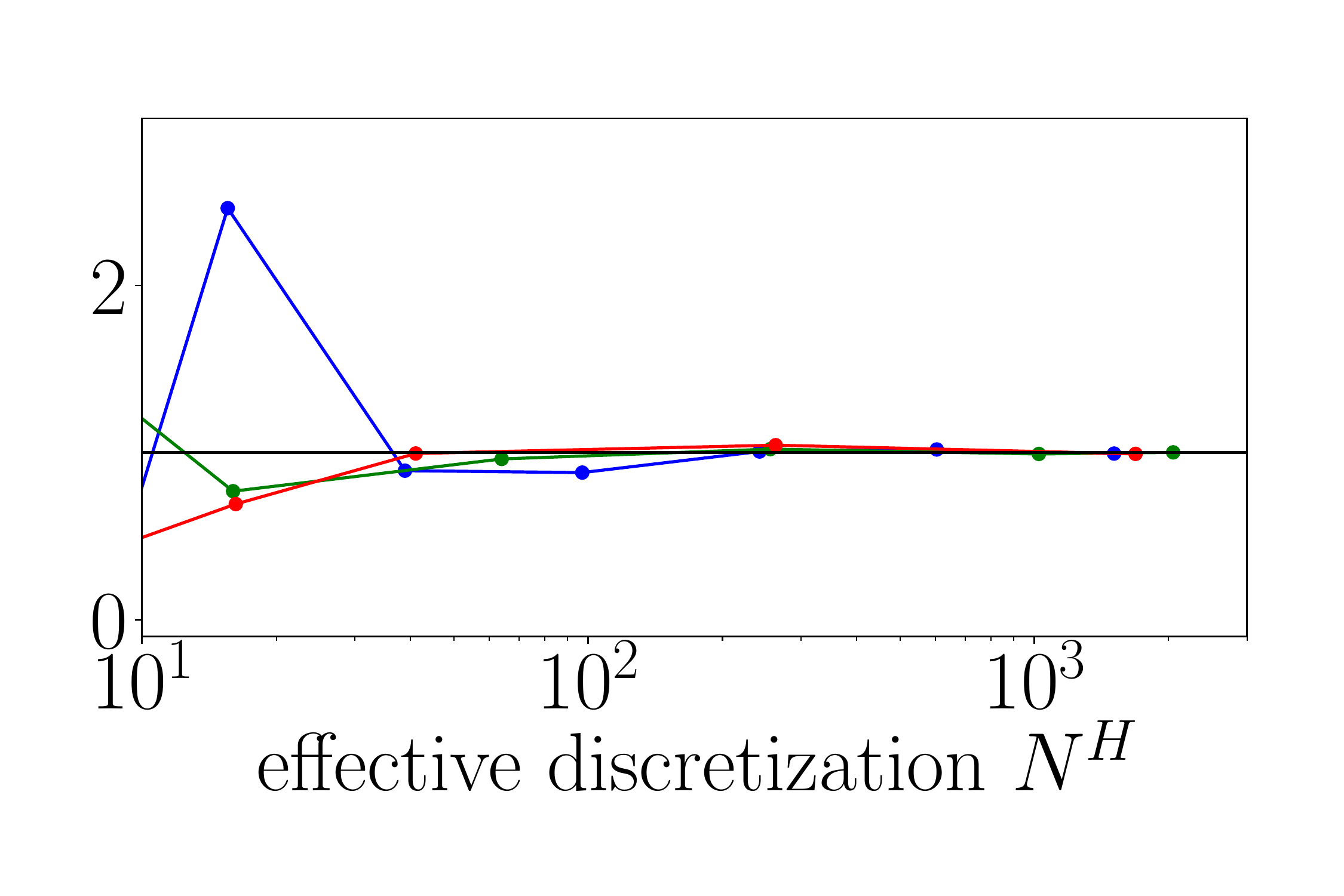}}}
\end{picture}}}
\caption{$(t_{\rm user}/{\rm iteration})^{1/3}$ plotted versus effective discretization $N^H$ for various values of $H$ (blue  $H=0.33$, green   $H=0.5$, red   $H=0.67$, cf.~Fig.~\ref{fig:performance}). They corroborate the estimate of ${\cal C}^{\rm ABSec} \sim \log(N_{\rm eff})^3$. Straight lines indicate fits of the form $ a \log(N)+b $. 
The inset shows the ratio between data points and the fit.}
\label{fig:complexity_vs_N}
\end{figure}

To estimate performance time, we observe that for values of $H \geq 0.33$, the number of additional gridpoints $M$ grows linearly in effective system size (cf.~Fig.~\ref{fig:MvsL}) throughout the entire observed range. Based on our empirical findings, we propose a linear relation $M \sim L = \log_2(N_{\rm eff})$, which implies, cf.~Eq.~(\ref{fig:MvsL}), an overall computational complexity of 
\be
{\cal C}^{\rm ABSec}(N_{\rm eff}) = {\cal O}\left(\left( \ln N_{\rm eff}\right)^{3}\right)\ , \qquad H \gtrsim \frac13\ ,
\label{eq:complexity_estimate}
\ee
since $M \gg 2^g$. This estimate is corroborated by Fig.~\ref{fig:complexity_vs_N}, where the scaling of user time with system size agrees with our estimate of $\log(N_{\rm eff})^3$ for sufficiently large system sizes.
The linear relation between the number of bisections $M$ and the logarithmic system size $L$, however, does not extend to smaller values of $H$, where Fig.~\ref{fig:MvsL}   indicates super-linear growth. Still, testing the ABSec routine at $H=0.25$ for an effective system size of $N_{\rm eff} = 2^{42}$ gave an average user time of $6.2s$ and was about $300 000$ faster than an extrapolation of the user time for DH at the same system size.\footnote{This experiment was run with an initial grid $\Lambda^4$ and $\varepsilon'=10^{-9}$.} This shows that for all practical purposes, ABSec remains a much faster algorithm even at parameters where estimate \eqref{eq:complexity_estimate} seems to no longer hold.

 For $H=0.33$, due to memory limitations,   DH is unable to generate paths larger than $N = 2^{24}$, where ABSec is already about 40 times faster. Since ABSec is also more memory-efficient (see next section), we can   generate  grids of   size up to $2^{32}$ for which, if we interpolate the growth of DH\footnote{Since DH scales with $N \log(N)$, we fit with $f(N;a,b,c) = N\left(a \log(N) + b  \right) + c$.}, we find that ABSec is 5500 times faster than DH for these parameters. For $H > \frac12$, the advantage is less pronounced, and at a comparable discretization precision, the algorithm is ``only'' 40-50 times faster at $H=0.67$.

\begin{figure}

\centering
\fboxsep0mm
\mbox{\includegraphics[trim=18 0 62 50,clip,width=1\columnwidth]{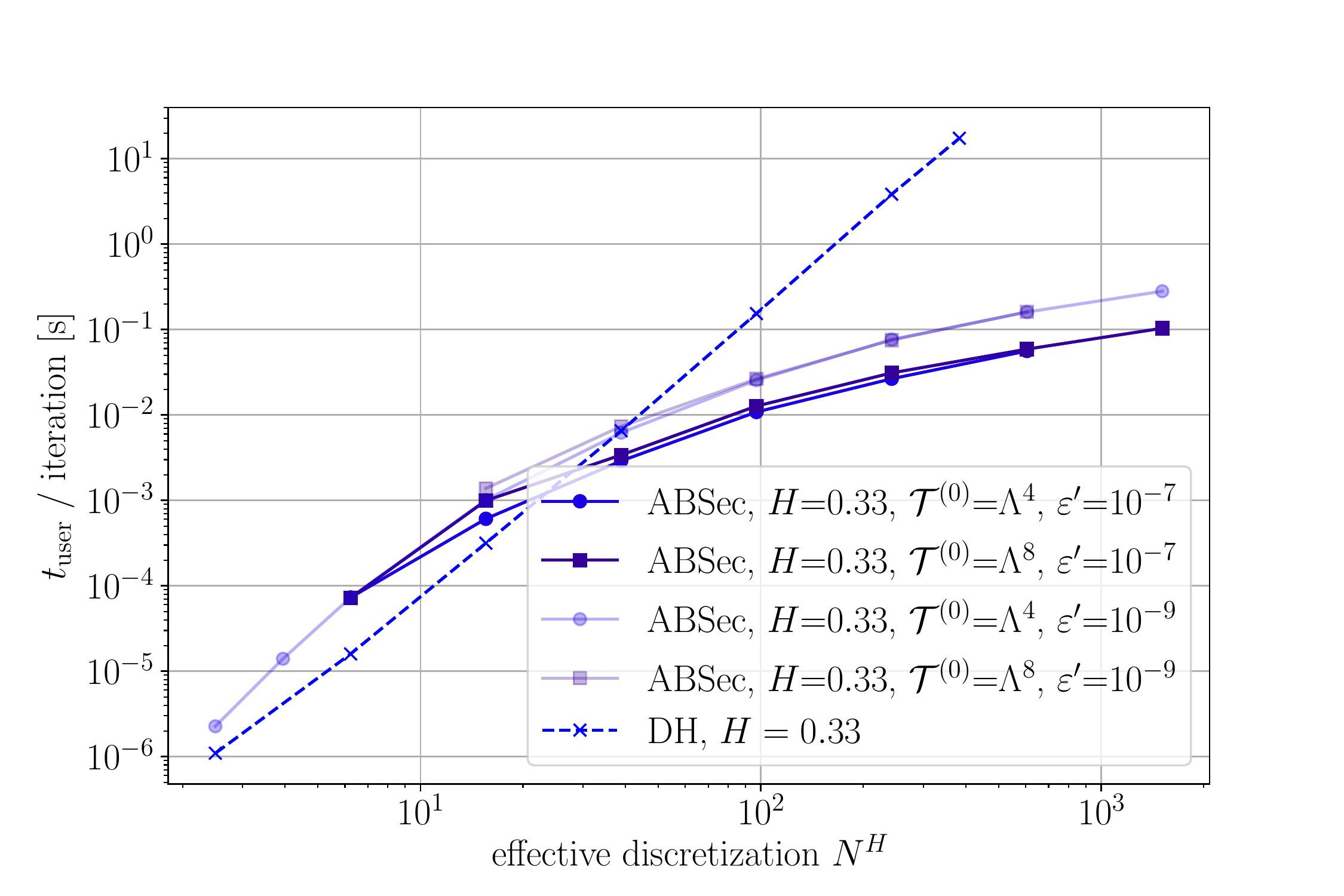}}
\caption{User time for ABSec compared to DH for two different initial grid sizes and two different values of error tolerance $\varepsilon'$. For a hundred times higher error tolerance, user time increase by up to $60 \%$.}
\label{fig:performance_vs_epsilon}
\end{figure}

 Performance also depends on the initial grid size. In Figs.~\ref{fig:performance} and \ref{fig:performance_vs_epsilon}, we compare run times for two different initial grids, $\Lambda^4$ and $\Lambda^8$. For larger initial grid sizes, the algorithm is slower since more points need to be generated initially. An increase in initial grid size   leads to a decrease of 15\% (for $H=0.33$)  in the average number of bisections. This is  approximately  outweighed by the time   DH takes to generate a path on $\Lambda^8$ (cf.~Fig.~\ref{fig:performance}).

The run time increases only slowly for a smaller error tolerance. In Fig.~\ref{fig:performance_vs_epsilon}, we show how user time decreases when changing $\epsilon'$ from $\varepsilon'= 10^{-9}$ to $\varepsilon'= 10^{-7}$. For an effective precision of $2^{\frac{32}{3}}$, user time increases by roughly 60 \%. Since error rates grow linearly with $\varepsilon'$ (see Fig.~\ref{fig:errorgrowth}), we conclude that for an error rate 100 times lower one only needs to invest 60\% more user time.

All together, these observations show that the algorithm behaves in a controlled manner for varying error tolerances and initial grid sizes. Depending on the number of iterations, and the quality of the data desired, choosing $g$ (initial grid size), $L$ (desired precision), and $ \varepsilon'$ (error tolerance level) accordingly  leads to an algorithm that  performs up to 5000 times faster than   DH at $H=0.33$, that was hitherto very hard to access with high precision. The algorithm should be tested more for $H=0.25$, where it allows one to reach a precision unimaginable by DH.

\subsection{Memory requirements}
\begin{figure}
\centering\fboxsep0mm
\mbox{\includegraphics[trim=22 10 63 50,clip,width=1\columnwidth]{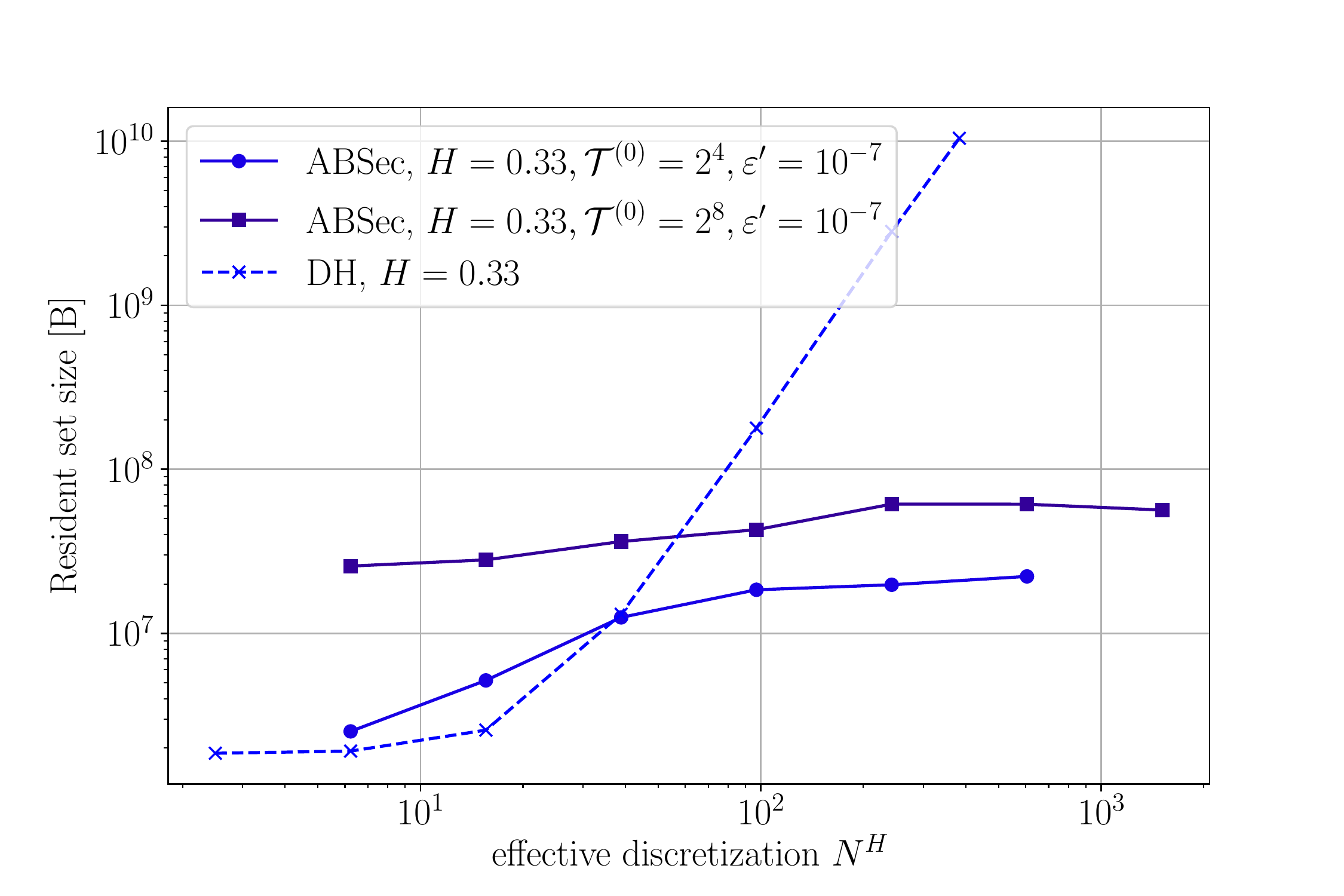}}
\caption{Memory usage for DH and ABSec for two different initial subgrid sizes. DH scales linearly in $N$, while ABSec grows only  slowly (see text for estimate). For system of size $N_{\rm eff} = 2^{28}$, ABSec needs only  $10^{-2}$ to $10^{-3}$ of the memory for DH. For larger systems or smaller $H$, the advantage of ABSec is even bigger. Measurements were taken after $10^4$ iterations.}
\label{fig:memory_vs_system_size}
\end{figure}

As a final  benchmark of our algorithm, we consider memory usage. The latter is defined by the resident set size of the process,  as measured by \texttt{getrusage}. When using DH, the full grid needs to be saved, and in  doing so memory usage scales like $N$. Fig.~\ref{fig:memory_vs_system_size} shows memory usage for both DH and ABSec when performed for different effective discretization precisions and initial grid sizes. It shows that for large system sizes, ABSec gains a growing and significant advantage. To generate a path of $2^{28}$ lattice points in double precision via  DH, one requires  10 GB working memory, whereas ABSec uses between 20 and 80 MB, depending on the initial grid size. This represents an  improvement by a factor of 125 to 500. This is due to the fact that only the initial grid which scales as $\cO(2^g)$, the additional gridpoints of order $\cO(M)$ and a correlation matrix, scaling as $\cO ( 2^g+M)^2 $, need to be stored. As implemented, additional memory is needed for the catalogue of inverse correlation matrices (cf. Eq.~\eqref{eq:CMatrixTable}) which occupies memory of order $\cO(2^{3g})$, so including the catalogue overall memory space grows like $2^{3g} + (2^g + M)^2$. Since we can assume that $2^g \ll M$, the necessary memory  grows with order $M^2$. For values of $H \gtrsim \frac13$, we empirically found that $M \sim \log( N_{\rm eff})$, such that in that parameter range we estimate memory to grow with
\begin{align}
{\cal M}^{\rm ABSec} (N_{\rm eff}) = \mathcal{O} \left( \log(N_{\rm eff})^2 \right)
 , \qquad H \gtrsim \frac{1}{3}\ .
\end{align}
 This advantage   is again due to  $ M \ll 2^L$, \ie using the fact that the first-passage time can be found to equal precision with much less grid points.

\subsection{Floating point precision}
\begin{figure}
\centering
\centerline{\fboxsep0mm
\mbox{\setlength{\unitlength}{1cm}\begin{picture}(8.65,5.6)
\put(0,0){\includegraphics[trim=10 10 53 40,clip,width=\columnwidth]{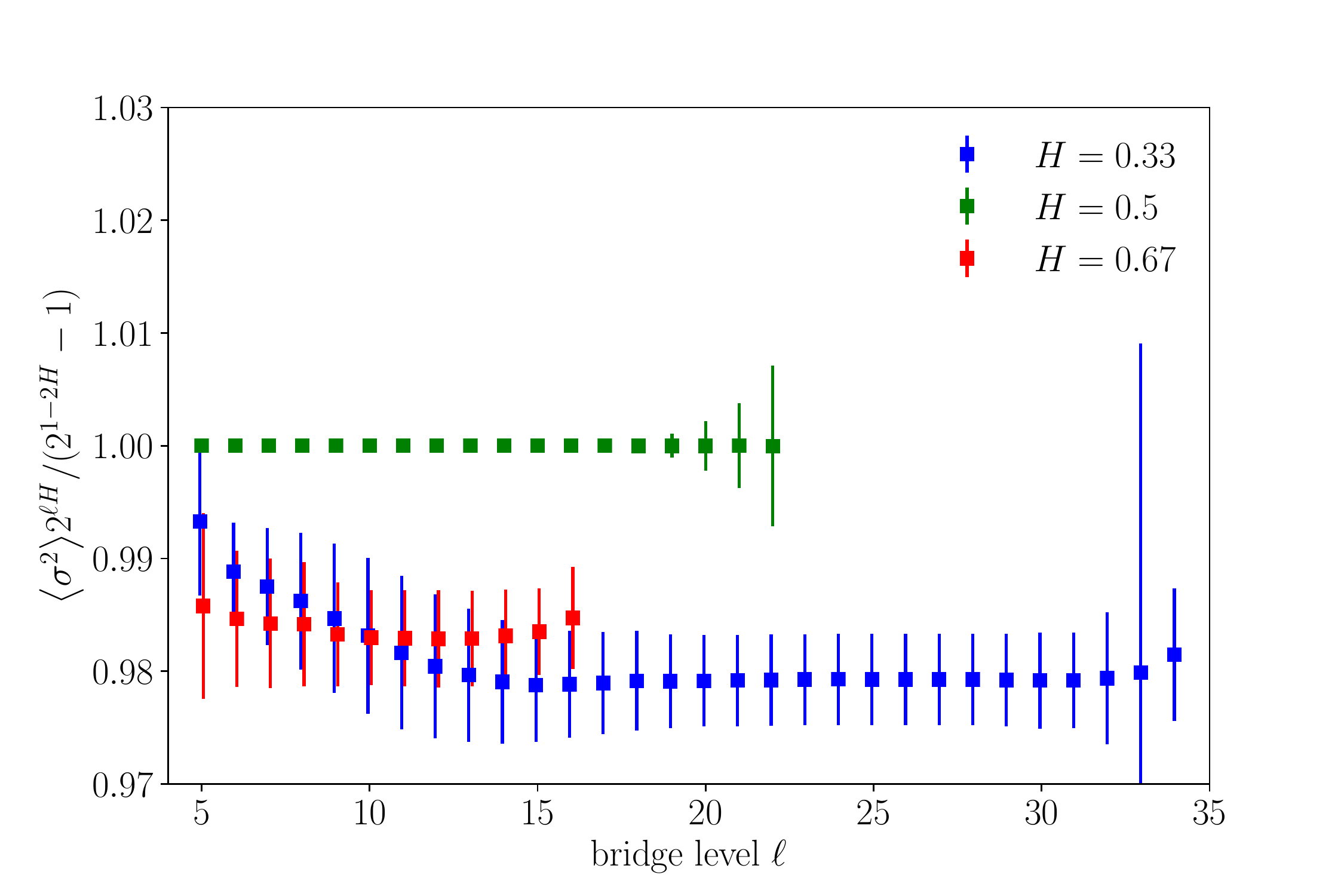}}
\put(1.3,3.24){\mbox{\includegraphics[trim=40 48 43 40,clip,width=0.42\columnwidth]{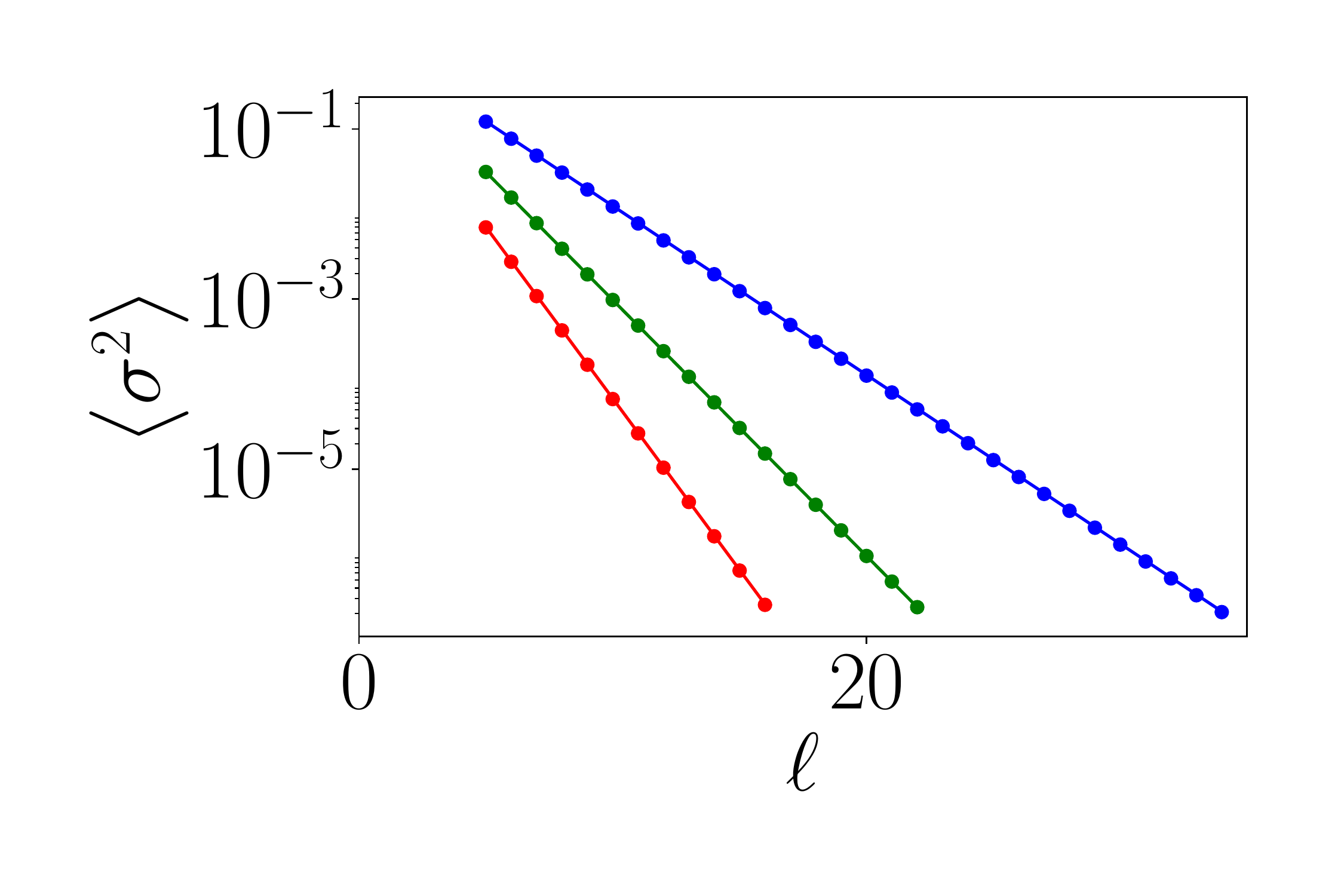}}}
\end{picture}}}
\caption{Ratio between sampled variance and no-neighbour-estimate of variance (cf.~Eq.~\eqref{eq:sigmanoneighbour}) of an inserted midpoint $X_{m}$ versus the level of the bisected bridge. For $H=0.5$ (green), the ratio equals 1, as  BM is Markovian. For $H\neq 0.5$, the variance fluctuates, as shown by the error bars  for one standard deviation.
Numerical errors due to a loss of floating point precision become relevant around $L_{\rm max} \simeq 11/H$. ABSec was used with an initial grid $\Lambda^8$ and $\varepsilon'=10^{-9}$.} 
\label{fig:variance_error}
\end{figure}

Currently, our implementation uses    the 64-bit \texttt{double} type. 
Since the variance of a bridge-point is calculated from the subtraction of   quantities of $\cO(1)$ (cf.~Eq.~\eqref{eq:sigma_from_grid}) whose difference can be as small as $\cO(2^{-LH})$, the subtraction suffers from the  finite  floating-point precision when $L$ is too large, as  is demonstrated in Fig.~\ref{fig:variance_error} (cf.~caption for details). This leads to  $L_{\rm max} \simeq 10.5/H$, or  $N_{\rm} \simeq 2\times 10^{\frac{3}{H}}$.

\subsection{Discussion}

In this section we   illuminated several aspects of our algorithm that show how it is capable of generating first-passage times with high numerical precision using several orders of magnitude less   CPU time and memory as compared to DH. We chose to compare ABSec to DH because the latter is widely spread in simulating first-passage times of fBm (see \eg \cite{Guerin2016,Levernier2018}), and since it is the fastest known exact generator of fBm. Since our method is also exact (the statistics of the grid generated is bias-free), we think of DH as the natural benchmark. There are     related approximative algorithms like the random midpoint displacement algorithm $R_{\ell, r}$ that also inserts midpoints, only taking into account the $\ell$ left and $r$ right nearest neighbours \cite{Norros2004}. This neglects long-range correlations between small increments at $t_1, t_2$ which even for  $t_1 \ll t_2$  are correlated algebraically via  $(t_1 - t_2)^{-1} + \cO\left( (H-\frac12)^2 \right)$ (for $H\neq \frac12$). The ABSec algorithm uses the full inverse correlation matrix of all points generated and is therefore closely related to exact procedures like DH.

Supported by our experiments, we are able to control both vertical and horizontal errors at the scale of inherent errors of a Monte Carlo simulation. In  practice, the limiting factors are not systematical errors of the algorithm but floating point imprecisions stemming from   the matrix inversion. 

The phone-book test used to asses the error rate does not take into account issues of precision when drawing new midpoints, which are copied from a pre-generated grid. Since this is an implementation-dependent grid, we decided to only use the phone-book test since the errors caused in that procedure are the ones inherent to the algorithm itself. An implementation with a higher-precision floating-point unit seems highly desirable.

\section{Summary}
\label{sec:summary}
When simulating first-passage times, or any other non-local observable, of fractional Brownian Motion, the large fluctuations for $H < \frac12$ require the grid to have a very high resolution for the same quality of data as for $H\geq\frac12$. Generating a fine grid is particularly expensive,   both in memory and time.
% for such highly self-correlated processes as fBm. 
The algorithm proposed here  refines the grid only where it is likely to impact the first-passage event. To give rigorous notion to that idea, we developed a precise criterion for when and where the grid should be refined. The  new mid-points are then sampled exactly. 
Comparing it to the fastest   known exact sampler, the Davies-Harte algorithm, we find that our implementation of the algorithm is   5000 times faster and uses  1000 times less memory when applied to $H=0.33$ at $N_{\rm eff}=2^{32}$, due to the fact that only roughly $0.1\%$ of the full grid is needed to determine the first-passage event. Our algorithm works with a probabilistic approximation, and the   error rate can be bounded by  $10^{-6}$ or even $10^{-8}$. This  should be sufficient for most Monte Carlo experiments and be in the order of numerical (algorithm-independent) errors. 

We have successfully used the algorithm to validate the    analytic results for  the first-passage time in Ref.~\cite{ArutkinWalterWiese2019}. There we used 2.5 CPU years at precision $N=2^{28}$. With DH we would have had to reduce the precision to $N=2^{24}$, which still would have taken 75 CPU years. 

Finally,   the concepts presented here can be  used for  other observables and other Gaussian processes. We   hope that our algorithm contributes to confirming theoretical predictions on extreme events in Gaussian processes that where hitherto   numerically inaccessible at the required precision.

 \section*{Acknowledgements} The authors   thank Marc-Thierry Jaekel and Andy Thomas for computing support and resources. B.W.~is grateful to Gunnar Pruessner for insightful discussions and support,  and thanks LPTENS and LPENS for hospitality. We thank Matteo D'Achille for a careful reading of the manuscript. 

\appendix

\section{Derivation of the critical strip length}
\label{app:criticalstrip}
In this section we derive the width of the critical strip which was introduced in Sec.~\ref{subsec:bridgecritical}.
The critical strip refers to the distance between a fBm-bridge of size $\delta t = 2^{-\ell}$ and the threshold $m$, below   which the midpoint of the bridge may surpass the threshold with probability larger than $\varepsilon$. We ignore any other grid points beyond the two fixed bridge points. By translational invariance, we    set $X_0 = 0$, and $X_{\delta t} = a$ ($a \in \mathbb{R})$. The problem  is then equivalently stated as 
\begin{align}
P(X^{\rm B}_{\delta t / 2} > c(\varepsilon) ) = \varepsilon,
\label{}
\end{align}
where $X^{\rm B}_t$ is the fBm-bridge process conditioned on $X_0, X_{\delta t}$. Following the derivation in \cite{DelormeWiese2016b}, the law of the fBm-bridge is itself   a Gaussian process with first and second moment,
\begin{align}
\avg{X^{\rm B}_t} &= \frac{\avg{X_t \delta(X_{\delta t} - a)}}{\avg{\delta(X_{\delta t} - a)}}\ , \\
\avg{X^{\rm B}_s X^{\rm B}_t} &= \frac{\avg{X_s X_t \delta(X_{\delta t} - a)}}{\avg{\delta(X_{\delta t} - a)}}\ ,
\label{}
\end{align}
where on the right-hand-side the averages are   over free fBm paths. As shown in Resf.~\cite{DelormeWiese2016b}, Eqs.~(8) and (9), the averages 
are
\begin{align}
\avg{X^{\rm B}_t} &= a \frac{C(t, \delta t)}{C(\delta t,\delta t)} \\
\avg{X^{\rm B}_s X^{\rm B}_t}^{\rm c} &= C(s, t) - \frac{C(s,\delta t)C(t,\delta t)}{C(\delta t,\delta t)} \ ,
\label{}
\end{align}
where $C(s,t)$ is the correlation function of Eq.~\eqref{eq:correlation_function}.
Since we are only interested in the midpoint with  $s=t=\delta t/2$, this yields \begin{align}
\mu = \avg{X_{\delta t/2}^{\rm B}} &= \frac{a}{2}\ , \\
\sigma = \avg{(X^{\rm B}_{\delta t/2})^2}^{\rm c} &=  \left( 2^{1-2H} -\frac12 \right) \left( \delta t \right)^{2H}\ .
\label{}
\end{align}
This determines the normal distribution of the midpoint and by translational invariance proves the values used in Sec.~\ref{subsec:bridgecritical}.

\section{How to generate an additional random   midpoint}

\label{app:derivationmidpoint}
We derive the conditional law of an additional  randomly generated midpoint for an  arbitrary Gaussian processes as given in Eqs.~\eqref{eq:conditionalsigma}--\eqref{eq:conditionalmu}. Let $\cT^N = t_1, \cdots, t_N$ and $\cX^N = X_{t_1}, \cdots, X_{t_N}$ be given, and denote the point to be inserted by $t_{N+1}$ and $X_{t_{N+1}}$ (the times are not  ordered). For ease of notation, we write $X_i = X_{t_i}$. As a Gaussian process, the vector $\vec{X} = (X_{1},\cdots,X_{N}, X_{{N+1}})^T$ is a normal random variable with mean zero and covariance matrix
\begin{align}
\avg{\vec{X} \vec{X}^T } = C(t_i, t_j) =: C(N+1)\ ,\quad 1\leq i,j \leq N\ .
\label{}
\end{align}
It has a symmetric inverse correlation matrix $C^{-1}_{i,j}$. Its probability law is therefore given by
\begin{align}
P(\vec{X}) = \frac{\exp\left( -\frac12 \sum_{i,j=1}^{N+1} X_i C^{-1}_{i,j} X_j\right)}{\sqrt{ (2 \pi)^{N+1} \det(C)   }}\ .
\label{B2}
\end{align}
Since $X_1, \cdots, X_N$ are fixed, $X_{{N+1}}$ conditioned on ${\cal X}^{N}$ follows the marginal distribution
\begin{align}
&P(X_{{N+1}} | \cX^N) \nonumber\\
&
=\frac{ \exp\left( -\frac12 X_{N+1}^2 C^{-1}_{N+1,N+1} - \sum_{j=1}^{N} X_j C^{-1}_{N+1, j} X_{N+1} \right)}{{\sqrt{2\pi/ C^{-1}_{N+1,N+1}} }}\ .
\label{B3}
\end{align}
Note that the normalizing factor in Eq.~\eqref{B2} has cancelled, since  Eq.~\eqref{B3} is a conditional average.  
This is a Gaussian distribution 
\begin{align}
&P(X_{{N+1}} | \cX^N) =  \frac{\exp\left( -\frac{\sigma^{2}}2  \left( X_{N+1} -\mu\right)^2 \right)}{ \sqrt{2\pi} \sigma}\ ,
\end{align}
with    variance
\begin{align}
\sigma^2 = \frac{1}{C^{-1}_{N+1, N+1}}\ ,
\end{align}
and mean
\begin{align}
\mu = - \sum_{j=1}^{N}X_j \frac{C^{-1}_{N+1,j}}{C^{-1}_{N+1,N+1}} .
\end{align}
The mean   can be seen as an average of the $X_{j}$ with   weight  $ {C^{-1}_{N+1,j}}/{C^{-1}_{N+1,N+1}}$.

\section{Derivation of the enlarged correlation matrix}
\label{app:derivationaugmentedmatrix}
In this section, we derive the algorithm to promote inverse correlation  matrices as given in Eqs.~\eqref{eq:defgammavec}--\eqref{eq:cmatrixpromoted}. Assuming that $C(N)$ and $C^{-1}(N)$ are  known, the aim is to find $C(N+1)$ and $C^{-1}(N+1)$ in as little as possible computational steps. The starting point is the observation that $C(N+1)$ contains $C(N)$ as block matrix and is only augmented by a row and identical column,
\begin{align}
C(N+1) =
\begin{pmatrix}
C(N)  &\vline & \vec{\gamma} \\
\hline 
\vec{\gamma}^T & \vline & \avg{X_{N+1}^2}
\end{pmatrix}
\label{eq:appcansatz}
\end{align}
where $\vec{\gamma}$ is defined in Eq.~\eqref{eq:defgammavec} and $\avg{X_{N+1}^2} = 2t_{N+1}^{2H}$ in the case of fBm, but is intentionally left general. 
For the more difficult part, the inversion, we assume that the inverse correlation matrix is of the form
\begin{align}
C^{-1}(N+1) = \begin{pmatrix}
A(N)  &\vline & \vec{b} \\
\hline 
\vec{b}^T & \vline & c
\end{pmatrix}
\label{eq:appcinvansatz}
\end{align}
for some arbitrary (symmetric) matrix $A$, vector $\vec{b}$ and number $c$. Multiplying matrices \eqref{eq:appcansatz} and \eqref{eq:appcinvansatz} results in
\begin{align}
C C^{-1} &=\begin{pmatrix}
C(N)A(N) + \gamma \otimes \vec{b}^T  &\vline & C(N)\vec{b}+c \vec{\gamma} \\
\hline 
(C(N)\vec{b}+c \vec{\gamma})^T & \vline & \vec{b}^T \vec{\gamma} + c \avg{X^2_{N+1}}
\end{pmatrix}\nonumber\\
& \stackrel != \mathbf{1}_{N+1}\ ,
\end{align}
such that one obtains the system of equations
\begin{align}
C(N)A(N)+\vec{\gamma} \otimes \vec{b}^T &= \mathbf{1}_N \ , \\
C(N) \vec{b} + c \vec{\gamma} &= \vec{0} \ , \\
\vec{b}^T \gamma + c \avg{X^2_{N+1}} &= 1\ .
\end{align}
This is solved by 
\begin{align}
A(N) &= C^{-1}(N)  + \frac{C^{-1}(N) \vec{\gamma}\otimes \vec{\gamma}^T C^{-1}(N)}{\avg{X_{N+1}^2}- \vec{\gamma}^T C^{-1} (N) \vec{\gamma} } \ ,\\
\vec{b} &=- \frac{C^{-1}(N) \vec{\gamma}}{\avg{X_{N+1}^2}- \vec{\gamma}^T C^{-1}(N) \vec{\gamma} } \ ,\\
c &= \frac{1}{\avg{X^2_{N+1}}-\vec{\gamma}^T C^{-1}(N) \vec{\gamma}}\ .
\end{align}
Defining $\vec{g}$ as in Eq.~\eqref{eq:defgvec} and $\sigma^2$ as in Eq.~\eqref{eq:sigma_from_grid}, one arrives at the inverse matrix \eqref{eq:cmatrixpromoted}.

\ifx\doi\undefined
\providecommand{\doi}[2]{\href{http://dx.doi.org/#1}{#2}}
\else
\renewcommand{\doi}[2]{\href{http://dx.doi.org/#1}{#2}}
\fi
\providecommand{\link}[2]{\href{#1}{#2}}
\providecommand{\arxiv}[1]{\href{http://arxiv.org/abs/#1}{#1}}
\providecommand{\hal}[1]{\href{https://hal.archives-ouvertes.fr/hal-#1}{hal-#1}}
\providecommand{\mrnumber}[1]{\href{https://mathscinet.ams.org/mathscinet/search/publdoc.html?pg1=MR&s1=#1&loc=fromreflist}{MR#1}}

\end{document}